\documentclass[a4paper,11pt]{article}

\usepackage[svgnames]{xcolor} 

\usepackage[printonlyused,withpage]{acronym} 
\usepackage{amsfonts}
\usepackage{anyfontsize}
\usepackage{bbm}
\usepackage{booktabs} 
\usepackage{braket}
\usepackage{cancel}
\usepackage{changepage}
\usepackage{enumitem}
\usepackage[mathscr]{euscript}
\usepackage[T1]{fontenc} 
\usepackage{gensymb}
\usepackage{Packages/jheppub}
\usepackage{listings} 
\usepackage{lmodern} 
\usepackage{makecell}
\usepackage{mathtools}
\usepackage[framemethod=tikz]{mdframed} 
\usepackage{microtype} 
\usepackage{multirow}
\usepackage{physics}
\usepackage{relsize}
\usepackage{rotating} 
\usepackage{sansmath}
\usepackage{simplewick}
\usepackage{slashed}
\usepackage{stmaryrd}
\usepackage{subcaption}
\usepackage{tcolorbox}
\usepackage{theorem}
\usepackage{tikz}
\usepackage[compat=1.1.0]{tikz-feynman}
\usepackage{contour}
\usepackage[normalem]{ulem}
\usepackage{xspace} 
\usepackage{yfonts}
\usepackage[vcentermath]{youngtab}

\tcbuselibrary{theorems}
\usetikzlibrary{decorations.markings}

\def\d{{\mathrm{d}}} 

\def\ii{{\text{i}}}

\makeatletter
\newcommand*{\transpose}{%
  {\mathpalette\@transpose{}}%
}
\newcommand*{\@transpose}[2]{%
  \raisebox{\depth}{$\m@th#1\intercal$}%
}
\makeatother

\newtcbox{\sln}{colback=Gainsboro,
colframe=Gainsboro}

\tikzset{snake it/.style={decorate, decoration={snake,amplitude=10mm}}}

\tikzset{/pgf/decoration/.cd,
    number of sines/.initial=10,
    angle step/.initial=20,
}

\newdimen\tmpdimen

\pgfdeclaredecoration{full sines}{initial}
{
    \state{initial}[
        width=+0pt,
        next state=move,
        persistent precomputation={
            \pgfmathparse{\pgfkeysvalueof{/pgf/decoration/angle step}}%
            \let\anglestep=\pgfmathresult%
            \let\currentangle=\pgfmathresult%
            \pgfmathsetlengthmacro{\pointsperanglestep}%
                {(\pgfdecoratedremainingdistance/\pgfkeysvalueof{/pgf/decoration/number of sines})/360*\anglestep}%
        }] {}
    \state{move}[width=+\pointsperanglestep, next state=draw]{
        \pgfpathmoveto{\pgfpointorigin}
    }
    \state{draw}[width=+\pointsperanglestep, switch if less than=1.25*\pointsperanglestep to final, 
        persistent postcomputation={
        \pgfmathparse{mod(\currentangle+\anglestep, 360)}%
        \let\currentangle=\pgfmathresult%
    }]{%
        \pgfmathsin{+\currentangle}%
        \tmpdimen=\pgfdecorationsegmentamplitude%
        \tmpdimen=\pgfmathresult\tmpdimen%
        \divide\tmpdimen by2\relax%
        \pgfpathlineto{\pgfqpoint{0pt}{\tmpdimen}}%
    }
    \state{final}{
        \ifdim\pgfdecoratedremainingdistance>0pt\relax
            \pgfpathlineto{\pgfpointdecoratedpathlast}
        \fi
   }
}

\pgfdeclaredecoration{completely sines}{initial}
{
    \state{initial}[
        width=+0pt,
        next state=upsine,
        persistent precomputation={\pgfmathsetmacro\matchinglength{
            \pgfdecoratedinputsegmentlength / int(\pgfdecoratedinputsegmentlength/\pgfdecorationsegmentlength)}
            \setlength{\pgfdecorationsegmentlength}{\matchinglength pt}
        }] {}
    \state{upsine}[width=\pgfdecorationsegmentlength,next state=downsine]{
        \pgfpathsine{\pgfpoint{0.25\pgfdecorationsegmentlength}{0.5\pgfdecorationsegmentamplitude}}
        \pgfpathcosine{\pgfpoint{0.25\pgfdecorationsegmentlength}{-0.5\pgfdecorationsegmentamplitude}}
    }
    \state{downsine}[width=\pgfdecorationsegmentlength,next state=upsine]{
        \pgfpathsine{\pgfpoint{0.25\pgfdecorationsegmentlength}{-0.5\pgfdecorationsegmentamplitude}}
        \pgfpathcosine{\pgfpoint{0.25\pgfdecorationsegmentlength}{0.5\pgfdecorationsegmentamplitude}}
}
    \state{final}{}
}

\tikzfeynmanset{ with arrow/.style = {
   decoration={
     markings,
     mark=at position 0.5
          with {\arrow[xshift=1.9mm]{Latex[width=2.75mm,length=3mm]}}
     },
   postaction=decorate}
}

\tikzfeynmanset{ with glarrow/.style = {
   decoration={
     markings,
     mark=at position 0.5
          with {\arrow[white,xshift=3mm]{Latex[width=4.125mm,length=4.5mm]}}
     },
   postaction=decorate}
}

\tikzfeynmanset{ with outarrow/.style = {
   decoration={
     markings,
     mark=at position 0.5
          with {\arrow[xshift=3.35mm]{Latex[width=4.58333mm,length=5mm]}}
     },
   postaction=decorate}
}

\tikzfeynmanset{ bigphoton/.style={
    /tikz/draw=none,
    /tikz/decoration={name=none},
    /tikz/postaction={
      /tikz/draw,
      /tikz/decoration={
        completely sines,
        segment length=3mm,
        amplitude=2.5mm,
      },
      /tikz/decorate=true,
    },
  },
}

\tikzfeynmanset{ roundbigphoton/.style={
    /tikz/draw=none,
    /tikz/decoration={name=none},
    /tikz/postaction={
      /tikz/draw,
      /tikz/decoration={
        full sines,
        segment length=3mm,
        amplitude=1.25mm,
      },
      /tikz/decorate=true,
    },
  },
}

\tikzset{ mega thick/.style= {line width = 3.4pt}
}

\hypersetup{colorlinks, breaklinks, linkcolor=black,citecolor=black,filecolor=black,urlcolor=black} 

\makeatletter
\renewcommand{\fnum@figure}{\textsc{\figurename~\thefigure}} 
\makeatother

\lstnewenvironment{pseudoc}{\lstset{frame=lines,language=C,mathescape=true}}{}
\lstnewenvironment{logs}{\lstset{frame=lines,basicstyle=\footnotesize\ttfamily,numbers=none}}{}
\lstnewenvironment{cc}{\lstset{frame=lines,language=C}}{}
\lstnewenvironment{c64}{\lstset{backgroundcolor=\color{c64},basewidth=0.65em,basicstyle=\commodoreface\color{c64light},numbers=none,framerule=10pt,rulecolor=\color{c64light},frame=tb,framexbottommargin=30pt}}{}
\lstnewenvironment{html}{\lstset{frame=lines,language=html,numbers=none}}{}
\lstnewenvironment{pseudo}{\lstset{frame=lines,mathescape=true,morekeywords={learn_string_domain, save_model}}}{}
\lstnewenvironment{pseudoctiny}{\lstset{language=C,mathescape=true,basicstyle=\tiny\sffamily}}{}
\lstnewenvironment{cctiny}{\lstset{language=C,basicstyle=\tiny\sffamily}}{}
\lstnewenvironment{pseudotiny}{\lstset{mathescape=true,basicstyle=\tiny\sffamily}}{}

\newcommand{\Lag}{\mathcal{L}}

\newcommand{\adj}{\ensuremath{{}^\dagger}}

\renewcommand{\bar}[1]{\ensuremath{\overline{#1}}}
\renewcommand{\tilde}[1]{\ensuremath{\widetilde{#1}}}

\title{Exploring the phenomenology of weak adjoint scalars in minimal \emph{R}-symmetric models}

\author{Linda M. Carpenter}
\author{and Matthew J. Smylie}
\affiliation{Department of Physics, The Ohio State University\\
191 W. Woodruff Ave,\\ Columbus, OH 43210}

\emailAdd{lmc@physics.osu.edu}
\emailAdd{smylie.8@osu.edu}

\date{\today}

\abstract{We examine the phenomenology of the scalar fields in weak and Higgs sectors of minimal $R$-symmetric models, in particular the `swino' and `sbino', the scalar partners to the chiral fields that marry the electroweak gauge bosons in Dirac gaugino models.  These fields are in adjoint representations of $SU(2)$ and $U(1)$ and have both $CP$-even and $CP$-odd components. The interactions of these new states are summarized, and decay widths are computed analytically to one loop order. We discuss the tree level contributions of these new states to the mass spectrum of MSSM sfermions. We also explore production cross sections and decay signatures at colliders for several chosen benchmarks. We find that large regions of parameter space are unconstrained by present collider data.}

\begin{document}
\maketitle

\section{Introduction}

Supersymmetry (SUSY) remains the leading framework for constructing theories beyond the Standard Model, due to its power in alleviating the hierarchy problem, producing viable dark matter candidates, and assisting gauge coupling unification at high energies. However, present collider data have not shown any novel supersymmetric phenomena, and parameter space of the simplest supersymmetric scenarios has been heavily constrained \cite{Aad:2020aze,Aad:2019byo}. Thus, there is a renewed interest in models that go beyond the Minimal Supersymmetric Standard Model (MSSM). One particularly intriguing class of models involves a continuous global $R$ symmetry, which forbids Majorana masses in favor of Dirac masses for the gauginos. Dirac gaugino models offer several features of theoretical and experimental interest; these include a natural hierarchy between gaugino and sfermion masses \cite{Fox:2002bu} and the suppression of sfermion production cross sections at colliders \cite{Kribs:2012gx}. In addition, these models can be interpreted as a promotion of the gauge and Higgs sectors to $\mathcal{N}=2$ supermultiplets \cite{Choi:2010dn,Itoyama:2011zi}, and they are compatible with general pictures of gauge mediated SUSY breaking \cite{Amigo:2009rsb,Carpenter:2010rsb,Carpenter:2017xru}.

Models with Dirac gauginos necessitate the existence of additional chiral fields  transforming in the adjoint representations of each Standard Model gauge group. The fermionic components of these fields `marry' to gauginos to produce Dirac gaugino masses. The scalar components of these chiral fields are known as the `sgauginos'.  These scalars have rich phenomenology and will be the focus of the present work. The minimal sgaugino sector consists of three complex scalars: one $SU(3)$ color octet denoted $O$, one $SU(2)$ triplet $T$,  which contains electrically charged and neutral components, and a fundamental scalar $S$. Like the Higgs, these scalar states have the same $R$-parity as the Standard Model fields; thus these models add an entirely new scalar sector to the SUSY spectrum.   

Recently, particular attention has been paid to the strong sectors of such models and the interactions of the color-octet \textit{sgluons} \cite{Plehn:2008ae,Kotlarski:2016lep,Diessner:2017sq,Carpenter:2020hyz,Carpenter:2020evo,Carpenter:2021vga}. Here, we turn our attention to the weak sector and examine the behavior of the scalar singlet and $SU(2)$ triplet fields.  Though the Higgs and electroweak-gaugino sectors of Dirac gaugino models have also been studied in detail in \cite{Choi:2010dn,Benakli_2011,Benakli:2013mdg,Goodsell_2020}, in-depth studies of the electroweak sgaugino sector have not yet been carried out. The sector consists of many new particles, the scalar and pseudoscalar singlet, scalar and pseudoscalar neutral triplet component, and the  charged scalar and pseudoscalar triplet components.  The neutral $CP$-even particles may participate in electroweak symmetry breaking and get vacuum expectation values (VEVs). In order to obey constraints from electroweak precision, neutral real triplet scalars must maintain very small VEVs.  Due to unavoidable interactions with the Higgs fields, these small VEVs mean that the $CP$-even components of the $SU(2)$ triplet state must be quite heavy---multi-TeV----in Dirac gaugino models. However, the $CP$-odd components may remain light. Therefore we consider here the low energy phenomenology of five states $S_R$, $S_I$, $T^0_I$ and $T^{\pm}_I$ where the indices $I$ and $R$ indicate $CP$-odd and -even components respectively.   

In this work, we create a complete compendium of the tree-level and one-loop level couplings of the $CP$-odd triplet and $CP$-even and -odd singlet states to MSSM fields. From this we compute the consequent decay widths, and calculate the decay spectra of branching fractions for several benchmark points in parameter space.  We find a markedly different decay pattern for scalar vs pseudoscalar states. Of particular importance are loop level couplings of electroweak sgauginos to pairs of SM gauge bosons. Such couplings allow for single production of neutral states through gluon fusion. This also allows di-boson resonant collider signatures through decay to gauge bosons---for example, a $W\gamma$ resonance decay of the charged triplet pseudoscalar. 

We compute the LHC loop production cross-sections of neutral sgauginos through gluon fusion and find interesting resonant effects that enhance the cross section due to sfermion mass thresholds.  Using these results along with computed branching fractions, we place bounds on parameter space from current LHC searches in a variety of final states including di-Higgs and di-tau final state searches.  We find that there is much parameter space for scalar neutral states under 1 TeV. We also compute production cross sections for pairs of charged and neutral triplet states through quark fusion and discuss relevant decay signatures.   

In this work we also study the effects of sgaugino interactions with Higgses on the MSSM mass spectrum.  In the most general $R$ symmetry preserving superpotentials, the $SU(2)$ and $U(1)$ adjoints may couple directly to the Higgs doublets.  This creates new contributions to the $\mu$ term, as in the $\mu$-less MSSM \cite{Nelson:2002ca}. We discuss how the neutral sgaugino VEVs directly impact the mass spectrum of MSSM sfermions, splitting masses within $SU(2)$ doublets beyond the limit due to $D$-terms in the MSSM, and we show that mass splittings between left and right handed sfermions are naturally induced.  We demonstrate how this may lead to SUSY mass spectra with sneutrino LSP/NLSP.

This paper will proceed as follows: first, we will describe the minimal model and its interactions. We will focus on the phenomenology of the real gauge singlet and the pseudoscalar triplet, providing their decay widths analytically to one loop order. Then, we will plot their branching fractions and production cross sections numerically for a range of benchmarks, comparing them to current constraints from collider data. Finally we will discuss prospects for searching for these particles.
\section{Model Summary}\label{model}

\subsection{Review of Dirac Gauginos}
Recall that in $D$-term implementations of Dirac gaugino models one may construct the Dirac gaugino masses by assuming a hidden sector $U(1)^{'}$ gauge field gets a $D$-term vev. We may thus write a new `supersoft' interaction \cite{Fox:2002bu} in the superpotential that couples the Standard Model field strength tensor to new chiral adjoint fields in the adjoint representations of the Standard Model gauge groups. Here we denote these fields as $A_i$, with the index $i$ running over the three gauge subgroups $SU(3),\ SU(2),$ and $U(1)$.

\begin{equation}
    W_{\mathrm{supersoft}}=\frac{\mathcal{W}^{'}_{\alpha}\mathcal{W}^{\alpha}_i A_i}{\Lambda_i}
\end{equation}
Here Lorentz indices are contracted between the hidden-sector and SM field strength tensors while the gauge indices are contracted between the SM field strength tensor and the adjoint chiral fields. The $\Lambda_i$ are the cutoff scales for the three operators. Once the hidden sector $D$ term is inserted, these operators become three independent Dirac mass terms for the SM gauginos:
\begin{equation}
    \frac{D}{\Lambda_i}\Psi_i\Psi_{Ai} \equiv m_{iD}\Psi_i\Psi_{Ai}
\end{equation}
The addition of the chiral adjoint fields to the theory introduces an entirely new sector to the theory, that  of the `sgauginos', the scalar components of the adjoint superfields. We normally denote the new chiral adjoint superfields as $S,T$, and $O$ for the $U(1)$, $SU(2)$, and $SU(3)$ adjoint fields respectively.  These must be complex fields, and therefore the sgaugino sector consists of several new independent spin zero states.  In this work, we will focus on the phenomenology of electroweak sgauginos $S$ and $T$, the `sbino' and `swino' states.

Beyond the supersoft operator written above, it is well known that a range of further SUSY breaking operators allow separate contributions to the scalar masses. One such operator is the `lemon-twist' term, which splits the masses of the real and imaginary components of the adjoints:

\begin{align}\label{eL}
W_{\mathrm{lemon-twist}}= \sum_{i=1}^3 \frac{\kappa_i}{\Lambda^2}\, \mathcal{W}'^{\alpha} \mathcal{W}_{\alpha}' A_i^a A_i^a.
\end{align} 

Previous work has studied the generation of this and other operators which lead to a viable phenomenological mass spectrum for scalar adjoint states \cite{Carpenter:2010rsb,Csaki:2013fla,Carpenter:2015mna}. For example, we may also consider the higher dimensional superpotential operators
\begin{equation}
    W_{\mathrm{trilinear}}=\zeta_S \frac{\mathcal{W}'\mathcal{W}'}{\Lambda^3}S^3 + \zeta_{TS}\frac{\mathcal{W}'\mathcal{W}'}{\Lambda^3} \Tr{TT}S,
\end{equation}
which contribute to the holomorphic scalar mass terms once vevs are inserted. We also refer the reader to \cite{Itoyama:2013sn,Itoyama:2013vxa} for discussion of dynamical $D$-term breaking and to \cite{Martin:2015eca} for non-supersoft methods of generating Dirac masses for gauginos.

\subsection{The electroweak scalar sector}

We will now examine the Higgs and electroweak sectors of a minimal Dirac gaugino model. The particle content of this model includes all the SM particles and their MSSM partners, as well as the additional gauge adjoints described above. The particles, with their charges under the SM gauge groups and the additional $R$ symmetry, are summarized in \hyperref[charges]{Table \ref{charges}}.

\begin{table}
\begin{center}
\begin{tabular}{|c|c|c|}
\hline
Supermultiplet & $SU(3)\times SU(2)_L\times U(1)_Y$ & $U(1)_R$ \\
\hline \hline
$(S,\ \psi_S)$ & $\boldsymbol{(1,1,0)}$ & 0 \\
\hline
$(T,\ \psi_T)$ & $\boldsymbol{(1,3,0)}$ & 0 \\
\hline
$(O,\ \psi_O)$ & $\boldsymbol{(8,1,0)}$ & 0 \\
\hline
$(H_u,\ \tilde{H}_u)$ & $\boldsymbol{(1,2,\frac{1}{2})}$ & +1 \\
\hline
$(H_d,\ \tilde{H}_d)$ & $\boldsymbol{(1,2,-\frac{1}{2})}$ & +1 \\
\hline
$(\tilde{q},\ q)$ & $\boldsymbol{(3,2,\frac{1}{6})}$ & $+\frac{1}{2}$ \\
\hline
$(\tilde{\bar{u}},\ \bar{u})$ & $\boldsymbol{(\bar{3},1,-\frac{2}{3})}$ & $+\frac{1}{2}$ \\
\hline
$(\tilde{\bar{d}},\ \bar{d})$ & $\boldsymbol{(\bar{3},1,\frac{1}{3})}$ & $+\frac{1}{2}$ \\
\hline
$(\tilde{l},\ l)$ & $\boldsymbol{(1,2,-\frac{1}{2})}$ & $+\frac{1}{2}$ \\
\hline
$(\tilde{\bar{e}},\ \bar{e})$ & $\boldsymbol{(1,1,1)}$ & $+\frac{1}{2}$ \\
\hline
\end{tabular}
\end{center}
\caption{The particle content of this model, along with gauge quantum numbers and the $R$-charge of the scalar part of the multiplet. These assignments of $R$-charge allow each term of the superpotential to have the necessary $R=2$. Here we use the two-component fermion notation of \cite{Martin:1997sp,Dreiner:2008tw}, in which $\bar{e} = e_R\adj$ and so on. Note that there is freedom to choose different $R$ for the squark and slepton fields, so long as the charges of corresponding left and right chiral fields add to 1.}
\label{charges}
\end{table}

The $R$-symmetry itself is a generalization of the discrete $R$-parity in the MSSM. However, care must be taken when considering the $\mu$ term in the MSSM superpotential. If all SM particles are assumed to have 0 $R$-charge, as is the case for MSSM $R$-parity, then the supersymmetric Higgs mass violates the symmetry. One approach to fixing this issue, used in fully $R$-symmetric models such as the Minimal $R$-Symmetric Standard Model (MRSSM)\cite{Kribs:2008rs,Diessner:2019sq}, requires the extension of the Higgs sector by additional $R$-Higgs fields. In this work, we will adopt the approach of more minimal models, such as the Minimal Dirac Gaugino Supersymmetric Standard Model (MDGSSM) \cite{Goodsell_2020, Chalons:2019md}, wherein the standard Higgses are charged under the $R$-symmetry. Electroweak symmetry breaking would then also break $R$ spontaneously, resulting in the generation of massless $R$-axions. To avoid this, there must be another source of $R$-breaking, typically written as an explicit $R$-breaking term in the soft Lagrangian. Of course, doing so will inevitably reintroduce a small Majorana mass for the gauginos, but we may consider this mass to be subdominant. The contributions to these Majorana gaugino masses will be at loop level and involve the soft $R$ breaking terms as well as Higgs VEV insertions. Since the Dirac mass terms can naturally be larger than the weak scale, we may approximate the Majorana mass as less than 10\% of the total mass. Thus, we will treat the gauginos as purely Dirac.

We augment the MSSM superpotential to include the supersoft interactions detailed above as well as additional trilinear terms allowed by symmetry:
\begin{equation}\label{superpot}
\begin{split}
W &= \left(\mu +\lambda_S S\right) H_u\cdot H_d + 2\lambda_T H_d\cdot TH_u +W_{\mathrm{supersoft}}+ W_{\mathrm{Yukawa}}, 
\end{split}
\end{equation}
where $S$ is the singlet, and $T$ is the triplet field expressed as
\begin{equation}
T = T^a \frac{\sigma^a}{2} = \frac{1}{2}\begin{pmatrix}
T^0 & \sqrt{2}T^+ \\
\sqrt{2}T^- & -T^0
\end{pmatrix},\ \ T^+ = \frac{T^1 -\ii T^2}{\sqrt{2}}, \ \ T^- = \frac{T^1 +\ii T^2}{\sqrt{2}}.
\end{equation}
The soft SUSY-breaking Lagrangian includes
\begin{equation}\label{soft}
\begin{split}
    -\Lag_{\mathrm{soft}} \supset\  & m^2_{H_u}|H_u|^2 + m^2_{H_d}|H_d|^2 + m^2_S |S|^2 + 2m^2_T\Tr(T\adj T)+\\ &+ \left(\frac{1}{2}B_S S^2 + B_T \Tr(TT) + h.c.\right) + LS + 
    \\
     &+ B_\mu H_u \cdot H_d + A_S S H_u\cdot H_d +A_T H_d \cdot TH_u + A_{SSS}S^3 + h.c.,
    \end{split} 
\end{equation}

In what follows, we will assume the only sources of R-breaking are those that aid  electroweak symmetry breaking and Higgs potential stability. As such, in general the $B_\mu$ term and linear term in $S$ are nonzero, and we will set all the trilinear $A$-terms to zero.
Integrating the supersoft terms in the above superpotential gives Dirac masses to the bino, the wino, and their scalar partners, and the resulting Lagrangian also includes trilinear scalar couplings:
\begin{equation}\label{trilinear}
\begin{split}
    -\Lag \supset\ & m_{1D}^2\left(S + S^{*}\right)^2 + m_{2D}^2\left(T^a + T^{*a}\right)^2 + \\ 
    & +\sqrt{2}gm_{2D}\left(T^a + T^{*a}\right)\sum_{j} \varphi^*_j\cdot \tau^a \varphi_j + \sqrt{2}g'm_{1D}\left(S + S^{*}\right)\sum_{j} \varphi^*_j Y_j\varphi_j,
\end{split}
\end{equation}
where the sums are over all scalar fields with nontrivial $SU(2)$ and $U(1)$ quantum numbers, respectively. We will further decompose the neutral states $S$ and $T^0$ into $CP$-even and $CP$-odd states:
\begin{equation}
S = \frac{v_S +S_R + \ii S_I}{\sqrt{2}},\ \  T^0 =\frac{v_T +T^0_R + \ii T^0_I}{\sqrt{2}}.
\end{equation}

The other two components of the swino, denoted $T^\pm$ above, remain charged under $U(1)_{EM}$ after electroweak symmetry breaking. Switching to a more convenient interaction basis, we adopt the notation
\begin{equation}
\begin{split}
    & T_R^+ = \frac{T^+ + (T^-)^*}{\sqrt{2}} = \frac{T_R^1 -\ii T_R^2}{\sqrt{2}},\ \  T_R^- = (T_R^+)^*, \\ & T_I^+ = \frac{T^+ - (T^-)^*}{\sqrt{2}} = \frac{\ii T_I^1 +T_I^2}{\sqrt{2}},\ \  T_I^- = (T_I^+)^*.
\end{split}
\end{equation}

There are multiple contributions to the masses of the real and imaginary parts of $S$ and $T$. In addition to non-holomorphic soft squared masses, $B$-terms, and supersoft masses for real components, there are additional contributions from $D$ term SUSY breaking operators. These operators include the `lemon-twist' term and the higher dimensional trilinear operators discussed above, which are of order $m_{iD}^2$ and split real and imaginary components of adjoints.  As such, when we specialize to the low energy phenomenology, we are free to treat the masses of the different components as independent parameters. We will denote the masses as \[M_{T^0_R}, M_{T^0_I}, M_{T^{\pm}_R}, M_{T^{\pm}_I}, M_{S_R}, M_{S_I} \] for the rest of this work.

The trilinear terms from the $D$ term interaction in Equation \ref{trilinear} present a crucial difference in the phenomenology of the $CP$-even and $CP$-odd states; the even states have access to tree level decays and interactions that the odd states do not. Analogously, only one of the charged triplet states couples to the $D$ term:
\begin{equation}
\begin{split}
    -\Lag &\supset 2g'm_{1D} S_R \sum_j Y_j |\varphi_j|^2 + 2gm_{2D} T^0_R \sum_j I_j |\varphi_j|^2 + \left(g m_{2D} T^+_R \sum_j \varphi_j\adj \sigma^+ \varphi_j + h.c.\right) \\
\end{split}
\end{equation}
The state $T^\pm_R$ then can decay to pairs of charged sfermions (for example here we have explicitly written couplings to third generation sfermions). Meanwhile, the ``odd'' state $T^\pm_I$ does not attain trilinear scalar couplings at tree level.

The trilinear terms in the superpotential also generate tree-level interactions between the scalars and higgsino-like neutralinos and charginos:
\begin{equation}
\begin{split}
	\Lag \supset &\frac{\lambda_T}{\sqrt{2}}\left(T_R^0 + \ii T_I^0\right)\left(\tilde{H}_u^0\tilde{H}_d^0 +\tilde{H}_d^-\tilde{H}_u^+\right)+ \frac{\lambda_S}{\sqrt{2}}\left(S_R+ \ii S_I\right)\left(\tilde{H}_u^0\tilde{H}_d^0 -\tilde{H}_d^-\tilde{H}_u^+\right)+ \\
	&+ T_R^-\left(\tilde{H}_d^0\tilde{H}_u^+ - \tilde{H}_d^+\tilde{H}_u^0 \adj\right)+ T_I^-\left(\tilde{H}_d^0\tilde{H}_u^+ + \tilde{H}_d^+\tilde{H}_u^0 \adj\right)  +h.c.
\end{split}
\end{equation}
In addition, there is the supersymmetric gauge interaction
\begin{equation}
\begin{split}
\Lag =& \sqrt{2}g (-\ii\epsilon_{abc})T^{*a}\tilde{W}^b\psi_T^c + h.c. \\
 &\supset g\left(T_R^0 +\ii T_I^0\right)\left(\tilde{W}^+\psi_T^- -\tilde{W}^-\psi_T^+\right) + h.c,
\end{split}
\end{equation}
so that the triplet scalars couple to the wino-like charginos.

\subsection{EWSB and mass eigenstates}

After electroweak symmetry breaking, the particle spectrum consists of four neutral $CP$-even scalars, three neutral $CP$-odd scalars, three independent charged scalars, six Majorana gauginos, and three charginos. It is convenient to work in the mass basis for all of these.

\subsubsection{Scalar Sector: Sgauginos and Higgs Fields}\label{mass-matrices}
Here we shall state the scalar mass matrices for this model. The full scalar potential is known, for a more complete discussion we refer to \cite{Choi:2010dn,Benakli_2011}.
Generically, all the neutral scalars will receive nonzero vacuum expectation values $v_u, v_d, v_S,$ and $v_T$. In addition, we may substitute $v_u$ and $v_d$ for the standard Higgs VEV $v= \sqrt{v_u^2 + v_d^2}$ and angle $\tan(\beta) = \frac{v_u}{v_d}$. The triplet VEV $v_T$ contributes to the $\rho$ parameter at tree level,
\begin{equation}
    \rho = 1 + 4\frac{v_T^2}{v^2},
\end{equation}
and is therefore constrained by electroweak precision measurements to be $\lesssim$ 3 GeV \cite{Benakli_2011}. Once VEVs are inserted, the $CP$-even scalars $S_R$ and $T^0_R$ will mix with the standard doublet Higgses. Given the superpotential in \hyperref[superpot]{Equation \ref{superpot}} and the soft terms of \hyperref[soft]{Equation \ref{soft}}, the real scalar mass terms can be written as $\Lag = \frac{1}{2} \Phi_0^{\mathrm{T}} M^2_H \Phi_0$, where $\Phi_0^{\mathrm{T}} = (h_d^0\ h_u^0\ S_R\ T_R^0)$, and the mass matrix itself can be written as
\begin{equation}\label{scalar-mass}
\begin{gathered}
     M_H^2 = 
    \begin{pmatrix}
    \mu_{\mathrm{eff}}^2 - \Delta m^2 + m_{H_d}^2 & \frac{1}{2}B_{\mu} & M^2_{13} & \frac{g}{2}m_{2D}v_d \\
   \frac{1}{2}B_{\mu} & \mu_{\mathrm{eff}}^2 + \Delta m^2 + m_{H_u}^2 & M^2_{23} & -\frac{g}{2}m_{2D}v_u \\
   M^2_{13} & M^2_{23} & M^2_{S_R} & \lambda_S \lambda_T v^2 \\
   \frac{g}{2}m_{2D}v_d & -\frac{g}{2}m_{2D}v_u &  \lambda_S \lambda_T v^2 & M^2_{T_R^0}
    \end{pmatrix}, \\ \\
     \mu_{\mathrm{eff}} = \mu + \lambda_S v_S + \lambda_T v_T, \\
     \Delta m^2 = g' m_{1D}v_S -g m_{2D}v_T, \\
     M^2_{13} = \lambda_S v_d(\mu + \lambda_T v_T)  - \frac{g'}{2}m_{1D}v_d, \\
     M^2_{23} = \lambda_S v_u(\mu + \lambda_T v_T) + \frac{g'}{2}m_{1D}v_u, \\
     M_{S_R}^2 = 2m_{1D}^2 + m_S^2 +B_S+ \lambda_S^2 v^2, \\ 
     M_{T^0_R}^2 = 2m_{2D}^2 + m^2_T +B_T+ \lambda_T^2 v^2.
\end{gathered}
\end{equation}
Here we have set the trilinear $A$-terms to zero, as previously noted. By exploiting the linear soft term, we are free to choose the scalar singlet VEV independently, and we take as large a triplet VEV as can be accommodated by precision measurements. We may then extract the physical scalar masses from the eigenvalues of the scalar matrix. In order to keep $v_T$ small, we may assume the triplet $T_R^0$ has a multi-TeV mass, and thus that mixing between $T_R^0$ and the other scalars is negligible. In general, the three remaining scalar states will mix nontrivially. For our purposes, we are most interested in scenarios with a clear hierarchy $M_{T_R^0} > m_{H^0}> M_{S_R} > m_h$, and where the heavier doublet mass is large enough to be considered decoupled from the low energy theory. This means that the largest mixing will be between the singlet and the lighter doublet, and we parameterize this mixing with a single angle $s_h = \sin(\theta_h)$. This does impact the phenomenology of the mostly-singlet state, as we will discuss in the next section, but we find it reasonable to treat the singlet $S_R$ as a mass eigenstate and $s_h$ as a perturbation. 

We also see from the trilinear scalar interactions in \hyperref[trilinear]{Equation \ref{trilinear}} that once VEVs are inserted, squarks and sleptons gain a contribution to their masses
\begin{equation}\label{mass_splitting}
\Delta m_j^2 = 2gm_{2D}v_T I_j + 2g'm_{1D}v_S Y_j,
\end{equation}
with $I_j$ the particle's weak isospin and $Y_j$ its hypercharge. The triplet VEV must be small, but if the Dirac wino mass $m_{2D}$ is in the multi-TeV range, this could be a significant source of mass splitting within weak doublets. This fact is particularly interesting, as it allows for a wider variety of possible particle spectra.

With A terms set to zero the $CP$-odd mass matrix may be expressed in the $(A^0,S_I,T_I^0)$ basis as
\begin{equation}
\begin{gathered}
    M_P^2 = 
    \begin{pmatrix}
    M_A^2 & 0 & 0 \\
    0 & M_{S_I}^2 & \frac{1}{2}\lambda_S \lambda_T v^2\\
    0 &  \frac{1}{2}\lambda_S \lambda_T v^2 & M_{T_I^0}^2
    \end{pmatrix}, \\ \\
    M_A^2 = \frac{2}{\sin(2\beta)}B_{\mu}, \\
    M_{S_I}^2 = m_S^2 -B_S + \lambda_S^2 v^2, \\
    M_{T^0_I}^2 = m_T^2 -B_T + \lambda_T^2 v^2.
\end{gathered}
\end{equation}
The off-diagonal term mixing the triplet and singlet is small compared to the diagonal. Thus we will take $(A^0,S_I,T_I^0)$ to be the approximate mass eigenbasis as well.

The charged swino states will mix with the charged Higgs as well as the Goldstone mode $a_G^\pm$. Thus we will write the charged mass matrix in the $(H^\pm,T_I^\pm,T_{RG}^\pm)$ basis, with the state
\begin{equation}
    T_{RG}^\pm = \frac{1}{\sqrt{\rho}}\left(T_R^\pm -2\frac{v_T}{v}a_G^\pm \right).
\end{equation}
We will also now assume that $\lambda_S,\ \lambda_T$ take their $\mathcal{N}=2$ SUSY values, which eliminates some of the charged scalar mixing terms. The mass matrix in this basis is then
\begin{equation}
\begin{gathered}
    M_C^2 =
    \begin{pmatrix}
    m^2_{H^+} & 0 & \sqrt{\rho}g m_{2D}v \sin(2\beta) \\
    0 & M_{T_I^\pm}^2 & 0 \\
    \sqrt{\rho}g m_{2D}v \sin(2\beta) & 0 & M_{T_R^\pm}^2 
    \end{pmatrix}, \\ \\
    m_{H^\pm}^2 = M_A^2 + M_W^2 + \frac{1}{2}(\lambda_T^2 - \lambda_S^2)v^2 + 2gm_{2D}\cos(2\beta)v_T + \\ + 2 \lambda_T^2 v_T^2 - 2\sqrt{2}\lambda_T v_T \mu_{\mathrm{eff}}, \\
    M_{T_I^\pm}^2 = M_{T_I^0}^2 + g^2v_T^2, \\
    M_{T_R^\pm}^2 = \rho M_{T_R^0}^2
\end{gathered}
\end{equation}
Note that the state $T_I^\pm$ does not mix significantly with the other states; also, at tree level there is a mass splitting between $T_I^\pm$ and $T_I^0$. This splitting is limited by the size of $v_T$, but it still may be of order GeV. This will have important phenomenological consequences, as we discuss in \hyperref[charged]{Section \ref{charged}}. The physical scalar masses and mixing parameters we select for our numerical analysis will be discussed in Section \ref{spectrum}.

\subsubsection{Fermion Sector: Gauginos and Higgsinos}
If the chargino mass matrix in the interaction basis is denoted $M_{Ch}$, then there exist two unitary matrices $U$ and $V$ such that $U^\mathrm{T} M_{Ch} V$ is diagonal. Then we can write
\begin{equation}
	\begin{pmatrix}
	\tilde{H}^-_d \\ \tilde{W}^- \\ \psi_T^-
	\end{pmatrix}
	= U
	\begin{pmatrix}
	\tilde{\chi}^-_1 \\ \tilde{\chi}^-_2 \\ \tilde{\chi}^-_3
	\end{pmatrix},
	\ \ \ \ 
	\begin{pmatrix}
	\tilde{H}^+_u \\ \tilde{W}^+ \\ \psi_T^+
	\end{pmatrix}
	= V
	\begin{pmatrix}
	\tilde{\chi}^+_1 \\ \tilde{\chi}^+_2 \\ \tilde{\chi}^+_3
	\end{pmatrix}.
\end{equation}
Similarly, there is a unitary $N$ such that $N^\mathrm{T}M_0 N$ diagonalizes the neutral mass matrix. Then
\begin{equation}
    \begin{pmatrix}
	\tilde{H}^0_u \\ \tilde{H}^0_d \\\tilde{W}^0 \\ \psi_T^0 \\\tilde{B}^0 \\ \psi_S
	\end{pmatrix}
	= N
	\begin{pmatrix}
	\tilde{\chi}^0_1 \\ \tilde{\chi}^0_2 \\ \tilde{\chi}^0_3 \\ \tilde{\chi}^0_4 \\ \tilde{\chi}^0_5 \\ \tilde{\chi}^0_6
	\end{pmatrix}.
\end{equation}
The neutralino and chargino mass spectra will in general have a profound effect on the observed phenomenology. We are primarily interested in the case where the LSP is a Higgsino-like neutralino, and the weak gaugino-like fermions are the heaviest. This is well motivated by naturalness considerations; the Dirac mass parameters of the supersoft operators can naturally be of order TeV or heavier, whereas the $\mu$ term is by necessity at the weak scale. Therefore, for the sake of simplicity, we will reduce the above fermion spectrum to a single Dirac chargino pair and a single pseudo-Dirac neutralino pair. That is, we assume that $\tilde{\chi}^0_1, \tilde{\chi}^0_2,$ and $\tilde{\chi}^\pm_1$ are approximately degenerate in mass, and we will assume that the wino- and bino-like fermions are heavy enough to be decoupled for our purposes. Explicitly, it is convenient to write the Higgsinos in four-component notation,
\begin{equation}
    \psi_H = \begin{pmatrix}
    \tilde{H}_u^0 \\ \tilde{H}_d^0 \adj
    \end{pmatrix},
    \ \ \ \ 
    \eta_H = \begin{pmatrix}
    \tilde{H}_u^+ \\ \tilde{H}_d^- \adj
    \end{pmatrix},
\end{equation}
and their gauge interactions as
\begin{equation}
\begin{split}
    \Lag \supset \left(\frac{g}{\sqrt{2}}\bar{\eta}_H \gamma_5 \gamma^{\mu} W_{\mu}^+ \psi_H + h.c.\right) -\frac{g}{2c_w} \bar{\psi}_H \gamma^{\mu}Z_{\mu} \psi_H +\\ +\frac{g}{c_w}\left(\frac{1}{2}-s_w^2\right)\bar{\eta}_H \gamma^{\mu}Z_{\mu} \eta_H + e \bar{\eta}_H \gamma^{\mu}A_{\mu} \eta_H,
\end{split}
\end{equation}
for ease of calculation.

In the fermion mass eigenbasis, we can summarize the above interaction terms with the following Lagrangian:
\begin{equation}
\begin{split}
	\Lag_{\phi\chi\chi} =& \mathcal{Y}^S_{ij}(S_R+\ii S_I) \tilde{\chi}^0_i\tilde{\chi}^0_j + \mathcal{Z}^S_{ij}(S_R+\ii S_I) \tilde{\chi}^+_i\tilde{\chi}^-_j + \\
	&+\mathcal{Y}^T_{ij}(T^0_R+\ii T^0_I) \tilde{\chi}^0_i\tilde{\chi}^0_j + \mathcal{Z}^T_{ij}(T^0_R+\ii T^0_I) \tilde{\chi}^+_i\tilde{\chi}^-_j + h.c,
\end{split}
\end{equation}
with the $\mathcal{Y}$ and $\mathcal{Z}$ matrices determined by the neutralino and chargino mixing parameters.
\begin{equation}
	\begin{split}
	&\mathcal{Y}^S_{ij} = \frac{\lambda_S}{\sqrt{2}}N_{ui}N_{dj}, \ \ \ \ \mathcal{Y}^T_{ij} = \frac{\lambda_T}{\sqrt{2}}N_{ui}N_{dj}, \\
	&\mathcal{Z}^S_{ij} = -\frac{\lambda_S}{\sqrt{2}}V_{ui}U_{dj}, \ \ \ \ \mathcal{Z}^T_{ij} = \frac{\lambda_T}{\sqrt{2}} V_{ui}U_{dj} + g\left(V_{Wi}U_{Tj}-V_{Ti}U_{Wj}\right).
	\end{split}
\end{equation}

Restricting to the Dirac Higgsinos described above, we can summarize all the tree-level interactions as
\begin{equation}\label{master}
\begin{split}
    \Lag_{\mathrm{int}} =\ & 2g'm_{1D} S_R \sum_j Y_j |\varphi_j|^2 +\frac{\lambda_S}{\sqrt{2}}S_R \left(\bar{\psi}_H \psi_H - \bar{\eta}_H \eta_H \right) + \\
    &+ 2gm_{2D} T^0_R \sum_j I_j |\varphi_j|^2 +\frac{\lambda_T}{\sqrt{2}}T^0_R \left(\bar{\psi}_H \psi_H + \bar{\eta}_H \eta_H \right) + \\
    &+ \frac{\ii\lambda_S}{\sqrt{2}} S_I\left(\bar{\psi}_H \gamma_5 \psi_H - \bar{\eta}_H \gamma_5 \eta_H \right) + \frac{\ii\lambda_T}{\sqrt{2}} T^0_I\left(\bar{\psi}_H \gamma_5 \psi_H + \bar{\eta}_H \gamma_5 \eta_H \right) + \\
    &+\left(T_R^+\bar{\eta}_H \gamma_5 \psi_H + T_I^+\bar{\eta}_H \psi_H + h.c.\right).
\end{split}
\end{equation}

\section{Extended scalar decay widths}
\subsection{Neutral Scalars}

Let us first examine the interactions and decay modes of the $CP$-even scalars. Because we assume the triplet scalar to be large compared to the weak scale, we will focus primarily on the real singlet $S_R$. Via the trilinear interactions from the $D$-term in \hyperref[trilinear]{Equation \ref{trilinear}}, $S_R$ can decay to a pair of any MSSM scalar with nonzero hypercharge. Thus the phenomenology will depend on the masses of all the MSSM scalars; if the lightest sfermions are far heavier than the Higgs, then the decay to two SM Higgses will be most relevant. Also important is the mixing angle $s_h$ introduced in \hyperref[mass-matrices]{Section \ref{mass-matrices}}.  As we will see later, heavy $S_R$ have no significant Higgs component, so below we have computed tree level decay widths of a pure singlet state. Lighter singlets have some measure of appreciable mixing with the Higgs. This  allows the singlet to decay via the tree-level processes of the Standard Model Higgs. This will enhance the partial width to top quarks and open decays to bottoms even if the sbottoms are decoupled.

In all, the tree-level partial widths can be summarized as:
\begin{align}
    \Gamma(S_R \rightarrow \varphi\varphi^*) &= \frac{Y_{\varphi}^2 g'^2}{8\pi}\frac{m_{1D}^2}{M_{S_R}}\sqrt{1-\frac{4m_{\varphi}^2}{M_{S_R}^2}} \\
    \Gamma(S_R \rightarrow hh) &= \frac{g'^2}{32\pi}\frac{m_{1D}^2}{M_{S_R}}(1-2s_h^2)^2\sqrt{1-\frac{4m_h^2}{M_{S_R}^2}} \\
    \Gamma(S_R \rightarrow \chi\bar{\chi}) &= \frac{\lambda_S^2}{16\pi}M_{S_R} \left(1 - \frac{4m_\chi^2}{M_{S_R}^2}\right)^{3/2}.
\end{align}

\begin{figure}
    \centering
 \begin{tikzpicture}[baseline={([yshift=-0.75ex]current bounding box.center)},xshift=12cm]
\begin{feynman}[large]
\vertex (i1);
\vertex [right = 1.25cm of i1] (i2);
\vertex [above right=1.5 cm of i2] (v1);
\vertex [below right=1.5cm of i2] (v2);
\vertex [right=1.25cm of v1] (g1);
\vertex [right=1.25cm of v2] (g2);
\diagram* {
(i1) -- [ultra thick, scalar] (i2),
(i2) -- [ultra thick, charged scalar] (v1),
(v1) -- [ultra thick, charged scalar] (v2),
(v2) -- [ultra thick, charged scalar] (i2),
(v1) -- [ultra thick, gluon, momentum=$k_1$] (g1),
(v2) -- [ultra thick, gluon, momentum'=$k_2$] (g2),
};
\end{feynman}
\node at (0.3,0.3) {$S_R$};
\node at (3.4,0.7) {$g^a$};
\node at (3.4,-0.65) {$g^b$};
\node at (1.5,0.8) {$\tilde{q}$};
\end{tikzpicture}
\ \ + \ \ 
\begin{tikzpicture}[baseline={([yshift=-.75ex]current bounding box.center)},xshift=12cm]
\begin{feynman}[large]
\vertex (i1);
\vertex [right = 1.25cm of i1] (i2);
\vertex [right= 1.25cm of i2] (g1);
\vertex [above right=1.5 cm of g1] (v1);
\vertex [below right=1.5cm of g1] (v2);
\diagram* {
(i1) -- [ultra thick, scalar] (i2),
(i2) -- [ultra thick, charged scalar, half left, looseness=1.7] (g1),
(g1) -- [ultra thick, charged scalar, half left, looseness=1.7] (i2),
(g1) -- [ultra thick, gluon,momentum={[arrow shorten=0.3]$k_1$}] (v1),
(g1) -- [ultra thick, gluon,momentum'={[arrow shorten=0.3]$k_2$}] (v2),
};
\end{feynman}
\node at (0.3,0.3) {$S_R$};
\node at (3.55,0.45) {$g^a$};
\node at (3.55,-0.375) {$g^b$};
\node at (1.875,1) {$\tilde{q}$};
\end{tikzpicture}
    \caption{The Feynman diagrams describing one-loop decay of the singlet to a gluon pair (up to permutation of the final state gluons). This and all subsequent diagrams were created with the aid of the Tikz-Feynman package \cite{Ellis:2017fd}.}
    \label{Sgg}
\end{figure}
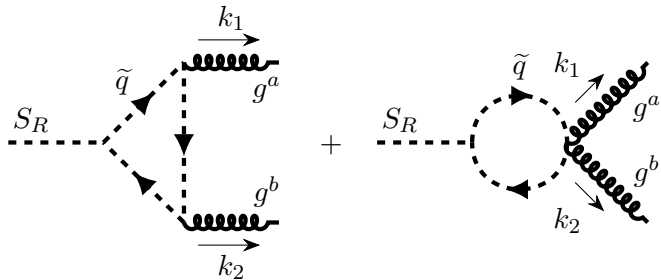

Relevant for LHC processes is the decay to gluons. This process occurs at one-loop level via the diagrams in \hyperref[Sgg]{Figure \ref{Sgg}}. For squarks of mass $m_j$ and hypercharge $Y_j$, the decay width can be expressed in terms of Passarino-Veltman integrals as
\begin{equation}\label{sgwidth}
    \Gamma(S_R \rightarrow gg) = \frac{g'^2 \alpha_s^2 m_{1D}^2}{16\pi^3 M_{S_R}} \left|\sum_j Y_j \left(1 + 32\ii\pi^2 m_j^2 C_0(M_{S_R}^2,0,0;m_j^2,m_j^2,m_j^2)\right) \right|^2.
\end{equation}
Here and for the rest of this work we use the convention
\begin{gather}
    B_0(k^2;s_1,s_2) = \int \frac{\dd^4 \ell}{(2\pi)^4} \frac{1}{(\ell^2-s_1)((\ell+k)^2-s_2)} \\
    C_0(k_1^2,k_1\cdot k_2, k_2^2; s_1,s_2,s_3) = \int \frac{\dd^4 \ell}{(2\pi)^4} \frac{1}{(\ell^2 - s_1)((\ell+k_1)^2-s_2)((\ell+k_2)^2-s_3)},
\end{gather}
and especially useful is the analytic form
\begin{equation}
\begin{split}
    C_0(M^2,0,0;m^2,m^2,m^2) &= \frac{\ii}{32\pi^2 M^2}\left(\log \frac{1+\sqrt{1-\tau}}{1-\sqrt{1-\tau}} \right)^2, \\
    \tau &= \frac{4m^2}{M^2}.
\end{split}
\end{equation}
This highlights an important difference from the scalar octet sector. The sgluon's decay widths feature a cancellation between the left- and right-chiral squarks; small stop mass splitting can result in long-lived states. Here, different squark hypercharges mean the cancellation is not exact.

Of course, the singlet can also decay to photons or weak bosons via similar one-loop processes to those above. Those diagrams are supplemented by loops of sleptons as well as squarks. There are realistic scenarios in which the squarks are much heavier than the sleptons. In these cases, the decay width to gluon pairs will be suppressed, and the branching fractions to $W^+W^-$ and to $\gamma\gamma$ will be more significant. In analogy with Equation \ref{sgwidth}, 
\begin{equation}
    \Gamma(S_R \rightarrow \gamma \gamma) = \frac{g'^2 \alpha_{EM}^2 m_{1D}^2}{16\pi^3 M_{S_R}} \left|\sum_j Y_j \left(1 + 32\ii\pi^2 m_j^2 C_0(M_{S_R}^2,0,0;m_j^2,m_j^2,m_j^2)\right) \right|^2.
\end{equation}
The analytic decay widths of $S_R$ to $W$ or $Z$ pairs do not simplify as easily due to their nonvanishing masses. Since they are rather cumbersome, we omit the expressions here.

\begin{figure}
    \centering
    \begin{tikzpicture}
	\begin{feynman}
	    \vertex (a) {\(S_R\)};
	    \vertex [right=of a](b);
	    \vertex [above right=of b](c);
        \vertex [below right=of b](d);
        \vertex [right=of c](f1){\(q\)};
        \vertex [right=of d](f2){\(q\)};
        
        \diagram*{
            (a) --[scalar] (b) --[scalar,edge label=\(\tilde{q}\)] (c) --[fermion] (f1),
            (b) --[scalar,edge label'=\(\tilde{q}\)](d) --[anti fermion] (f2),
            (c) --[plain,gluon, edge label=\(\tilde{g}\)] (d),
        };
	\end{feynman}
	\end{tikzpicture}
    \caption{The one-loop decay of $S_R$ to a quark pair.}
    \label{Srqq}
\end{figure}
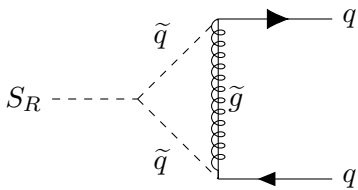

There is additionally a loop-level decay to quarks, with the dominant contribution shown in Figure \ref{Srqq}. This decay is enhanced by mixing with the Higgs doublet state, which has the Standard Model Yukawa interactions at tree level. The decay width is
\begin{equation}
    \begin{split}
    \Gamma(S_R \rightarrow q\bar{q}) =& \frac{3}{2\pi}m_q^2 M_{S_R} \left|\frac{s_h}{v}+ \sum_j Y_j \mathcal{I}_{qj}\right|^2\left(1-\frac{4m_q^2}{M_{S_R}^2}\right)^{3/2}, \\
    & \mathcal{I}_{qj}=\frac{2g'\alpha_s}{3\pi}\frac{m_{1D}}{M_{S_R}^2-4m_q^2}16\ii\pi^2\left(B_0(m_q^2; m_{D3}^2, m^2_{\Tilde{q}_j}) - B_0(M_{S_R}^2; m^2_{\Tilde{q}_j}, 
   m^2_{\Tilde{q}_j}) +\right. \\
   & \ \ \ \left. - (m_q^2 + m_{3D}^2 - m_{\Tilde{q}_j}^2) C_0(m_q^2, \frac{1}{2}M_{S_R}^2 - m_{\Tilde{q}_j}^2, m_q^2; m^2_{D3}, m^2_{\Tilde{q}_j}, m^2_{\Tilde{q}_j})\right).
    \end{split}
\end{equation}
In the limit where the singlet is light, the mixing with the SM Higgs is large, and that term is significant. If the singlet is heavy, then the mixing is negligible compared to the loop term. This is only relevant for third generation quarks, and effectively it causes the branching fraction to $b\bar{b}$ to dominate if $S_R$ is too light to decay on shell.

The benchmarks we consider assume a very heavy triplet scalar, but for completeness we will outline its tree-level decay widths here:
\begin{align}
    \Gamma(T^0_R \rightarrow \varphi\varphi^*) &= \frac{I_{\varphi}^2 g^2}{8\pi}\frac{m_{2D}^2}{M_{T_R^0}}\sqrt{1-\frac{4m_{\varphi}^2}{M_{T_R^0}^2}} \\
    \Gamma(T^0_R \rightarrow hh) &= \frac{g^2}{32\pi}\frac{m_{2D}^2}{M_{T_R^0}}\sqrt{1-\frac{4m_h^2}{M_{T_R^0}^2}} \\
    \Gamma(T^0_R \rightarrow \chi\bar{\chi}) &= \frac{\lambda_T^2}{16\pi}M_{T_R^0} \left(1 - \frac{4m_\chi^2}{M_{T_R^0}^2}\right)^{3/2}.
\end{align}
\subsection{Pseudoscalars}

We shall now discuss the decay modes of the pseudoscalar adjoints. Note that unlike the $CP$-even states, these do not have trilinear couplings generated from the supersymmetric $D$-term. Thus, the only tree-level interactions they have are via the Higgs couplings of the superpotential and, in the case of the triplet, the $SU(2)$ interaction. In the scenario described above, with a single Dirac neutralino and a single Dirac chargino, both of which are entirely Higgsino-like, the tree-level interactions can be simplified to:
\begin{equation}
    \Lag_{\mathrm{int}} = \frac{\ii \lambda_S}{\sqrt{2}} S_I (\bar{\psi}_H \gamma_5 \psi_H - \bar{\eta}_H \gamma_5 \eta_H) +\frac{\ii \lambda_T}{\sqrt{2}} T^0_I (\bar{\psi}_H \gamma_5 \psi_H + \bar{\eta}_H \gamma_5 \eta_H)
\end{equation}

We will narrow our focus to the triplet state $T^0_I$. The only tree-level decay available to it is to two Higgsinos, with width
\begin{equation}
    \Gamma(T_I^0 \rightarrow \chi\bar{\chi}) = \frac{\lambda_T^2}{16\pi}M_{T_I^0} \sqrt{1-\frac{4m_{\chi}^2}{M_{T_I^0}^2}}.
\end{equation}
If this state is too light to decay to two LSPs, loop processes will be important. In the case of the pseudoscalar $SU(3)$ octet, the decay to two top quarks is a significant one-loop channel \cite{Darme:2018rec}, and the same decay is available to the weak pseudoscalar, as seen in Figure \ref{TIloop} below.

\begin{figure}[h]
	\centering
	\begin{tikzpicture}
	\begin{feynman}
	    \vertex (a) {\(T_I^0\)};
	    \vertex [right=of a](b);
	    \vertex [above right=of b](c);
        \vertex [below right=of b](d);
        \vertex [right=of c](f1){\(\gamma / Z\)};
        \vertex [right=of d](f2){\(\gamma / Z\)};
        
        \diagram*{
            (a) --[scalar] (b) --[plain,boson] (c) --[boson] (f1),
            (b) --[plain, boson](d) --[boson] (f2),
            (c) --[plain, boson, edge label=\(\tilde{\chi}^+\)] (d),
        };
	\end{feynman}
	\end{tikzpicture} \ \ 
	\begin{tikzpicture}
	\begin{feynman}
	    \vertex (a) {\(T_I^0\)};
	    \vertex [right=of a](b);
	    \vertex [above right=of b](c);
        \vertex [below right=of b](d);
        \vertex [right=of c](f1){\(Z\)};
        \vertex [right=of d](f2){\(Z\)};
        
        \diagram*{
            (a) --[scalar] (b) --[plain,boson,edge label=\(\tilde{\chi}^0\)] (c) --[boson] (f1),
            (b) --[plain, boson,edge label'=\(\tilde{\chi}^0\)](d) --[boson] (f2),
            (c) --[plain, boson, edge label=\(\tilde{\chi}^0\)] (d),
        };
	\end{feynman}
	\end{tikzpicture} \\ \vspace{0.5cm}
	
	\begin{tikzpicture}
	\begin{feynman}
	    \vertex (a) {\(T_I^0\)};
	    \vertex [right=of a](b);
	    \vertex [above right=of b](c);
        \vertex [below right=of b](d);
        \vertex [right=of c](f1){\(W^+\)};
        \vertex [right=of d](f2){\(W^-\)};
        
        \diagram*{
            (a) --[scalar] (b) --[plain,boson,edge label=\(\tilde{\chi}^+\)] (c) --[boson] (f1),
            (b) --[plain, boson,edge label'=\(\tilde{\chi}^-\)](d) --[boson] (f2),
            (c) --[plain, boson, edge label=\(\tilde{\chi}^0\)] (d),
        };
	\end{feynman}
	\end{tikzpicture} \ \  
	\begin{tikzpicture}
	\begin{feynman}
	    \vertex (a) {\(T_I^+\)};
	    \vertex [right=of a](b);
	    \vertex [above right=of b](c);
        \vertex [below right=of b](d);
        \vertex [right=of c](f1){\(t\)};
        \vertex [right=of d](f2){\(t\)};
        
        \diagram*{
            (a) --[scalar] (b) --[plain,boson,edge label=\(\tilde{\chi}^0\)] (c) --[fermion] (f1),
            (b) --[plain, boson,edge label'=\(\tilde{\chi}^0\)](d) --[anti fermion] (f2),
            (c) --[scalar, edge label=\(\tilde{t}\)] (d),
        };
	\end{feynman}
	\end{tikzpicture}
	\caption{The one-loop decays of the neutral pseudoscalar triplet.}
	\label{TIloop}
\end{figure}
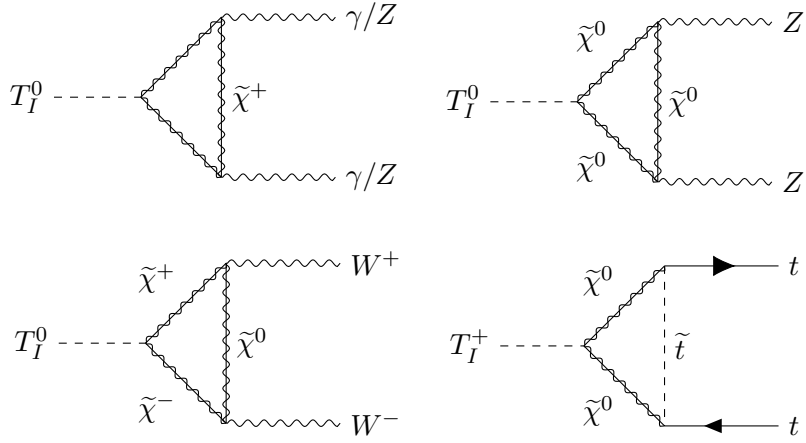

This decay can be expressed analytically as:
\begin{equation}
\begin{split}
    \Gamma(T^0_I \rightarrow t \bar{t})=\ &\frac{\lambda_T^2 y_t^4}{16\pi} m_{\chi}^2 m_t^2 M_{T_I^0} \left(1 - \frac{4m_t^2}{M_{T_I^0}^2}\right)^{3/2} |C_L - C_R|^2, \\
    & C_j = C_0(m_t^2, m_t^2, M_{T_I^0}^2, m_{\chi}^2, m_{\chi}^2, m_{\tilde{t}j}^2)
    \end{split}
\end{equation}

It is known that, for group theory reasons, the decay of the odd sgluon to two gluons vanishes at one loop \cite{Plehn:2008ae}. However, there is no such reason that loop decays of the triplet or singlet to bosons (see Figure \ref{TIloop}) would cancel. The branching fractions to electroweak bosons then become dominant below the threshold for two LSPs.

In fact, we can compute the decay rate of the pseudoscalars to two photons to be
\begin{equation}
	\Gamma(T^0_I\rightarrow \gamma\gamma)= 16\pi \lambda_T^2 \alpha_{EM}^2 M_{T_I^0}^3 m_{\chi}^2 \left| C_0(0,0,M_{T_I^0}^2;m_{\chi}^2,m_{\chi}^2,m_{\chi}^2)\right|^2.
\end{equation}
Similarly, the pseudoscalar can decay to one photon and one $Z$, with partial width
\begin{equation}
\begin{split}
	\Gamma(T^0_I\rightarrow Z\gamma)=& \lambda_T^2 \alpha_{EM} \frac{g^2}{c_W^2}\left(\frac{1}{2}-s_W^2\right)^2 M_{T_I^0}^3 m_{\chi}^2\ \times \\
	\times &\left| C_0(0,M_Z^2,M_{T_I^0}^2;m_{\chi}^2,m_{\chi}^2,m_{\chi}^2)\right|^2 \left(1-\frac{M_Z^2}{M_{T_I^0}^2}\right)^3.
\end{split}
\end{equation}
The decay to two $Z$ or $W$ bosons will also involve a neutralino loop.
\begin{equation}
\begin{split}
	\Gamma(T^0_I \rightarrow ZZ)=& \frac{\lambda_T^2 g^4}{4\pi c_W^4} M_{T_I^0}^3 m_{\chi}^2 \left|\left(\frac{1}{2}-s_W^2\right)^2 + \frac{1}{4}\right|^2 \times \\
	&\times \left| C_0(M_Z^2,M_Z^2,M_{T_I^0}^2;m_{\chi}^2,m_{\chi}^2,m_{\chi}^2)\right|^2 \left(1-\frac{4M_Z^2}{M_{T_I^0}^2}\right)^{3/2}.
\end{split}
\end{equation}
\vspace{0.5cm}
\begin{equation}
\begin{split}
    \Gamma(& T^0_I \rightarrow W^+ W^-)= \frac{\lambda_T^2 g^4}{4\pi} M_{T_I^0}^3 m_{\chi}^2 \left|\mathcal{I}_{WW}(M_{T_I^0}) \right|^2 \left(1-\frac{4M_W^2}{M_{T_I^0}^2}\right)^{3/2}, \\
    &\mathcal{I}_{WW}(M_{T_I^0}) = \frac{1}{M_{T_I^0}^2 - 4M_W^2}\left[ M_{T_I^0}^2 C_0(M_W^2,M_W^2,M_{T_I^0}^2;m_{\chi}^2,m_{\chi}^2,m_{\chi}^2)+ \right. \\ 
     & \ \ \ \ \  \left. +4\left(B_0(M_{T_I^0}^2,m_{\chi}^2,m_{\chi}^2)-B_0(M_W^2,m_{\chi}^2,m_{\chi}^2)\right) \right].
\end{split}
\end{equation}

The only difference between the singlet and the triplet is in their couplings to the Higgsinos. Pseudoscalar singlet widths can be expressed as
\begin{gather}
    \Gamma(S_I \rightarrow \chi\bar{\chi}) = \frac{\lambda_S^2}{16\pi}M_{S_I} \sqrt{1-\frac{4m_{\chi}^2}{M_{S_I}^2}} \\
	\Gamma(S_I\rightarrow \gamma\gamma)= 16\pi \lambda_S^2 \alpha_{EM}^2 M_{S_I}^3 m_{\chi}^2 \left| C_0(0,0,M_{S_I}^2;m_{\chi}^2,m_{\chi}^2,m_{\chi}^2)\right|^2 \\[1.5ex]
	\begin{split}
	    \Gamma(S_I\rightarrow Z\gamma)= & \lambda_S^2 \alpha_{EM} \frac{g^2}{c_W^2}\left(\frac{1}{2}-s_W^2\right)^2 M_{S_I}^3 m_{\chi}^2\times \\
	    & \times\left| C_0(0,M_Z^2,M_{S_I}^2;m_{\chi}^2,m_{\chi}^2,m_{\chi}^2)\right|^2 \left(1-\frac{M_Z^2}{M_{S_I}^2}\right)^3
	\end{split}\\[1.5ex]
	\begin{split}
	    \Gamma(S_I \rightarrow ZZ)= & \frac{\lambda_S^2 g^4}{4\pi c_W^4} M_{S_I}^3 m_{\chi}^2 \left|\left(\frac{1}{2}-s_W^2\right)^2 - \frac{1}{4}\right|^2\times \\
	    & \times\left| C_0(M_Z^2,M_Z^2,M_{S_I}^2;m_{\chi}^2,m_{\chi}^2,m_{\chi}^2)\right|^2 \left(1-\frac{4M_Z^2}{M_{S_I}^2}\right)^{3/2}
	\end{split}\\[1.5ex]
	\Gamma(S_I \rightarrow W^+ W^-)= \frac{\lambda_S^2 g^4}{4\pi} M_{S_I}^3 m_{\chi}^2 \left| \mathcal{I}_{WW}(M_{S_I}) \right|^2 \left(1-\frac{4M_W^2}{M_{S_I}^2}\right)^{3/2} \\
	\Gamma(S_I \rightarrow t \bar{t})=\ \frac{\lambda_S^2 y_t^4}{16\pi} m_{\chi}^2 m_t^2 M_{S_I} \left(1 - \frac{4m_t^2}{M_{S_I}^2}\right)^{3/2} |C_L - C_R|^2
\end{gather}

Neither of the pseudoscalars couple to gluons at one loop, and the decay rates to quarks are proportional to the quark mass. As a result, hadron colliders may be ill-suited for the single production of these particles.
\subsection{Charged Scalars}\label{charged}

We will now outline the decay modes of the charged adjoint scalars. As discussed in Section \ref{model}, the two charged swino states behave similarly to the neutral scalar and pseudoscalar states; one has trilinear couplings to sfermions and one does not. As such, much of the previous two sections applies equally well to the charged scalars. The state $T^\pm_R$ will decay at tree level to two sfermions if heavy enough. For eg. one stop and one sbottom, the width is given by
\begin{equation}
\begin{split}
    \Gamma(T_R^+ &\rightarrow \Tilde{t}\ \Tilde{b}\adj) = \frac{g^2}{32\pi} \frac{m_{2D}^2 \sqrt{\lambda(M_{T_R^\pm}^2,m_{\tilde{t}}^2,m_{\tilde{b}}^2)}}{M_{T_R^\pm}^3}, \\
    &\lambda(x,y,z) = x^2 +y^2 +z^2 -2xy-2yz-2zx
\end{split}
\end{equation}
and similar for other weak doublet scalars. Its tree-level decay width to higgsinos is given by
\begin{equation}
    \Gamma(T_R^+ \rightarrow \tilde{\chi}^+\tilde{\chi}^0) = \frac{\lambda_T^2}{16\pi}M_{T_R^\pm} \sqrt{1-\frac{4m_{\chi}^2}{M_{T_R^\pm}^2}}.
\end{equation}
We expect that this state is relatively heavy, so we will focus our attention on the ``odd'' state $T_I^\pm$. Its only tree-level interactions are the gauge coupling and the coupling to Higgsinos; this decay width is given by
\begin{equation}
    \Gamma(T_I^+ \rightarrow \tilde{\chi}^+\tilde{\chi}^0) = \frac{\lambda_T^2}{16\pi}M_{T_I^\pm} \left(1 - \frac{4m_\chi^2}{M_{T_I^\pm}^2}\right)^{3/2}
\end{equation}
If too light to decay to supersymmetric particles, its dominant decay modes will be to $W^\pm \gamma$, $W^\pm Z$, or quark pairs at the one-loop level, as shown in \hyperref[Tchvv]{Figure \ref{Tchvv}}. The $W^\pm \gamma$ channel in particular is an interesting signature, and we feel it is worth examining more closely in the future.

\begin{equation}
\begin{split}
    \Gamma(& T^+_I \rightarrow W^+ \gamma)= \frac{\lambda_T^2 g^4}{4\pi} M_{T_I^\pm}^3 m_{\chi}^2 \left|\mathcal{I}_{W\gamma} \right|^2 \left(1-\frac{M_W^2}{M_{T_I^\pm}^2}\right)^{3/2}, \\
    &\mathcal{I}_{W\gamma} = C_0(0,M_W^2,M_{T_I^\pm}^2;m_{\chi}^2,m_{\chi}^2,m_{\chi}^2)+ \frac{2\left(B_0(M_{T_I^\pm}^2,m_{\chi}^2,m_{\chi}^2)-B_0(M_W^2,m_{\chi}^2,m_{\chi}^2)\right)}{M_{T_I^\pm}^2 - M_W^2}.
\end{split}
\end{equation}
The decay rate to a top/bottom quark pair is given by
\begin{equation}
    \Gamma(T^+_I \rightarrow t \bar{b})= \frac{\lambda_T^2 y_t^4}{32\pi} m_{\chi}^2 m_t^2 M_{T_I^\pm} \left(1 - \frac{m_t^2}{M_{T_I^\pm}^2}\right)^{2} |C_L - C_R|^2,
\end{equation}
where we have taken $m_b \ll m_t$, and $C_L$ and $C_R$ are defined in the previous section.

\begin{figure}[h]
	\centering
	\begin{tikzpicture}
	\begin{feynman}
	    \vertex (a) {\(T_I^+\)};
	    \vertex [right=of a](b);
	    \vertex [above right=of b](c);
        \vertex [below right=of b](d);
        \vertex [right=of c](f1){\(W^+\)};
        \vertex [right=of d](f2){\(\gamma / Z\)};
        
        \diagram*{
            (a) --[scalar] (b) --[plain,boson,edge label=\(\tilde{\chi}^0\)] (c) --[boson] (f1),
            (b) --[plain, boson,edge label'=\(\tilde{\chi}^+\)](d) --[boson] (f2),
            (c) --[plain, boson, edge label=\(\tilde{\chi}^+\)] (d),
        };
	\end{feynman}
	\end{tikzpicture} \ \ \ \ 
	\begin{tikzpicture}
	\begin{feynman}
	    \vertex (a) {\(T_I^+\)};
	    \vertex [right=of a](b);
	    \vertex [above right=of b](c);
        \vertex [below right=of b](d);
        \vertex [right=of c](f1){\(t\)};
        \vertex [right=of d](f2){\(b\)};
        
        \diagram*{
            (a) --[scalar] (b) --[plain,boson,edge label=\(\tilde{\chi}^0\)] (c) --[fermion] (f1),
            (b) --[plain, boson,edge label'=\(\tilde{\chi}^+\)](d) --[anti fermion] (f2),
            (c) --[scalar, edge label=\(\tilde{t}\)] (d),
        };
	\end{feynman}
	\end{tikzpicture}
	\caption{The one-loop decays of the charged triplet.}
	\label{Tchvv}
\end{figure}
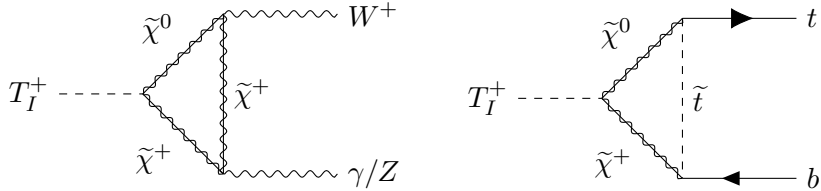

\begin{figure}[h]
    \centering
    \begin{tikzpicture}
    \begin{feynman}
        \vertex (a) {\(T_I^+\)};
        \vertex [right=of a](b);
        \vertex [above right=of b](f1){\(T_I^0\)};
        \vertex [below right=of b](c);
        \vertex [above right=of c](f2){\(f\)};
        \vertex [below right=of c](f3){\(f'\)};
        
        \diagram*{
            (a) --[scalar] (b) --[scalar] (f1),
            (b) --[boson,edge label'=\(W^+\)] (c),
            (c) --[fermion] (f2),
            (c) --[anti fermion] (f3),
        };
    \end{feynman}
    \end{tikzpicture}
    \caption{The three body decay of the charged scalar.}
    \label{3body}
\end{figure}
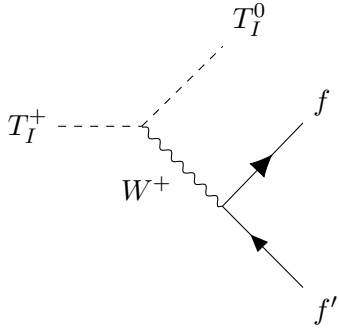

In addition, a mass splitting between the neutral pseudoscalar $T_I^0$ and the charged $T_I^+$ allows a three body decay, as shown in \hyperref[3body]{Figure \ref{3body}}. The splitting will be $\order{1\ \mathrm{GeV}}$, so the $W$ boson in the diagram will be far off shell, and only leptons and light flavor quarks will be decay products. We may then write the partial decay width as
\begin{equation}
\begin{split}
    \Gamma(T_I^+ \rightarrow & T_I^0 f \bar{f}') = \frac{g^4}{8\pi^3} \frac{M_{T_I^\pm}}{M_W^4} \int_0^{(M_{T_I^\pm} - M_{T_I^0})^2} \mathrm{d}s \ \mathcal{F}(s), \\ \\ 
    & \mathcal{F}(s) = \frac{s}{2M^2_{T_I^\pm}}\sqrt{\lambda(M^2_{T_I^\pm},M^2_{T_I^0},s)} + \frac{s^2}{2(M^2_{T_I^\pm}-M^2_{T_I^0} + s)} \log\left(\frac{1-\Delta(s)}{1+\Delta(s)} \right), \\ \\
    & \Delta (s) = \frac{(M^2_{T_I^\pm}-M^2_{T_I^0} + s)\sqrt{\lambda(M^2_{T_I^\pm},M^2_{T_I^0},s)}}{\lambda(M^2_{T_I^\pm},M^2_{T_I^0},s) + 2s M^2_{T_I^\pm}},
\end{split}
\end{equation}
where $M_{T_I^\pm}$ and $M_{T_I^0}$ are the masses of the charged and neutral scalar respectively, and the triangular function $\lambda(x,y,z)$ is defined above. This decay can result in signatures with emission of a single soft lepton.

\section{Mass Spectrum and Benchmarks}\label{spectrum}

We will compute the above decay widths numerically in order to connect with present and future experiments. First we must outline a mass spectrum and set of benchmarks to use. As noted in \hyperref[mass-matrices]{Section \ref{mass-matrices}}, we suppose $H^0, H^{\pm}$, and $A$ are heavy enough to be considered decoupled. Due to trilinear couplings with the Higgs, the field $T_R^0$ must get a VEV. In order to suppress this VEV below a few GeV, the mass $T_R^0$ must lie in the multi-TeV range, therefore this field is effectively decoupled from light states. We will first discuss the issue of sgauginos mixing with the states in the Higgs sector.

The mass matrix in \hyperref[scalar-mass]{Equation \ref{scalar-mass}} will have a state with a significant singlet-scalar component. We can vary the mass of the singlet-like state by varying the diagonal mass term $M_{S_R}^2$. The mixing parameter $s_h$ is not independent; once Dirac mass, mu term and coupling parameters are fixed, this mixing shifts with the $S_R$ mass.  It is small for heavy singlets and increases as the singlet mass approaches the SM Higgs mass from above. For our numerical analysis below, we compute the mass eigenvalue and mixing parameter for several values of the soft singlet mass and then interpolate. Specific values are given in \hyperref[shmix]{Table \ref{shmix}} for our benchmark parameters detailed below. This mixing only affects the singlet branching fractions for relatively light singlet masses.  In this case there is a non-negligible Higgs component which enhances decays into third generation quarks when tree level decays to gauginos and scalar particles are kinematically inaccessible. Larger mixing angles will be disfavored by Higgs sector measurements. Indeed, as the weak triplet is effectively decoupled, the $CP$-even scalar sector of this model closely resembles that of the Next-to-Minimal Supersymmetric Standard Model (NMSSM). Mixing in the Higgs sector of the NMSSM has been well studied, and LHC Higgs measurements favor those scenarios where the 125 GeV state is mostly doublet-like \cite{Jeong:2014xaa,Carena:2015moc, Choi:2019yrv}. Explicitly, our angle $\theta_h$ is referred to as $\theta_2$ in the referenced literature, and our benchmarks assume the angle $\theta_1 \sim 0$. Then the constraints from Higgs mixing in \cite{Choi:2019yrv} translate to the constraint $|\sin\theta_h|\lesssim 0.4$. For our chosen parameters, this translates to $M_{S_R}\geq 250$ GeV.

\begin{table}[]
    \centering
    \begin{tabular}{|c|c|}
        \hline
        $M_S$ (GeV) & $s_h$ \\
        \hline
        198 & -0.503 \\
        \hline
        206 & -0.450 \\
        \hline
        269 & -0.220 \\
        \hline
        306 & -0.161 \\
        \hline
        371 & 0.102 \\
        \hline
        423 & 0.0759 \\
        \hline
        512 & 0.0501 \\
        \hline
        587 & 0.0374 \\
        \hline
    \end{tabular}
    \caption{Selected values of the singlet mass and mixing with the lightest Higgs.}
    \label{shmix}
\end{table}

The $CP$-odd scalar sector will have three physical states: the MSSM pseudoscalar $A^0$ and the neutral adjoints $S_I$ and $T_I^0$.  It is worth noting that the CP-odd sgauginos do not have trilinear supersoft couplings to Higgs fields. These fields do not gain VEVs, and their masses may be taken as independent of other model parameters. In our numerical analysis below, we assume the MSSM-like $A^0$ remains heavy compared to the SM Higgs and decouples in the case of zero $A$-like terms, and we study the case where either $S_I^0$ or $T_I^0$ is taken to be light. Therefore the remaining light mass eigenstate will approximately be pure $S_I$ or $T_I^0$.  As this model contains possibly light, $CP$-odd states that decay to bosons, certain scenarios could in principle allow for axion or axion-like particle searches \cite{Graham:2015ouw,Florez:2021zoo} to be relevant, though we do not consider any such scenarios here. As noted previously, the pseudoscalar adjoints do not couple to gluons at one loop, so the production cross section is dominated by quark fusion in the case of $T_I^0$ or vector boson fusion/associated production in the case of $S_I$.

We assume that the third generation squarks and sleptons are the lightest sparticles, and the others may be considered to decouple from the theory. As explained previously, we are most interested in the case where the lightest supersymmetric particles are the higgsino-like neutralinos and charginos, followed by the left handed sleptons, the sneutrino and the staus. Within each generation of sfermions, mass splittings arise due to trilinear couplings with electroweak sgauginos, referenced in Equation \ref{mass_splitting}. Additional mass contributions for one generation of sleptons are
\begin{equation}
\begin{split}
    & \Delta m^2_{\Tilde{\tau},L} =  -gm_{2D}v_T- g' m_{1D} v_S, \ \ \ \Delta m^2_{\Tilde{\nu},L} =  gm_{2D}v_T- g' m_{1D} v_S, \\ & \Delta m^2_{\Tilde{\tau},R} =  2 g' m_{1D} v_S,
\end{split}
\end{equation}
while for each generation of up- and down-type squarks 
\begin{equation}
\begin{split}
    \ \Delta m^2_{\Tilde{t}L} =  gm_{2D}v_T+ \frac{1}{3}g' m_{1D} v_S, \ \ \ &\Delta m^2_{\Tilde{b}L} = - gm_{2D}v_T + \frac{1}{3}g' m_{1D} v_S,  \\
  \ \ \   \Delta m^2_{\Tilde{t}R} =  -\frac{4}{3} g' m_{1D} v_S, \ \ \  &\Delta m^2_{\Tilde{b}R} =  \frac{2}{3} g' m_{1D} v_S.
\end{split}
\end{equation}

The singlet VEV adds to mass splitting between the left and right handed states with each generation. The triplet VEV creates mass splittings between the upper and lower components of the isospin doublet beyond contributions from the standard model $D$-term. If the triplet VEV is at its limit of a few GeV, with Dirac masses are multi-TeV, this splitting is $\order{10-50\ \mathrm{GeV}}$ between states of the same weak doublet. A phase choice should allow arbitrary sign choice of these terms.

As stated previously, the triplet VEV contributes to the electroweak $\rho$ parameter at tree level and is thus constrained to be a few GeV or smaller. Other corrections to this parameter will in general occur at loop level and will be proportional to $\lambda_S$ or $\lambda_T$, the only model parameters which break the custodial symmetry. In models such as the $\mu$-less MSSM \cite{Nelson:2002ca}, the dominant contribution to $W$ and $Z$ self-energies comes from loops involving triplet fermions; here we assume these states are married to the heavy wino and bino and are effectively decoupled. Therefore the only surviving corrections are due to loops entirely composed of $CP$-odd scalars. By dimensional analysis these corrections are already smaller than those from light triplet fermions. Further, the net contribution to $\rho$ must be due to the mass splitting between $T_I^0$ and $T_I^\pm$, which is also controlled by the VEV $v_T$. Similarly, contributions to the effective operators $|HD_\mu H|^2$ and $[H\adj W^{\mu\nu} H]B_{\mu\nu}$ are suppressed by loops containing heavy particles.

With all of this in mind, we outline multiple benchmark scenarios in \hyperref[benchmarks]{Table \ref{benchmarks}}. For benchmarks B1 and B2, stop masses are chosen agreeing with the most optimistic scenario in \cite{Carpenter:2020hyz}. The sneutrino mass is relatively less constrained, and viable scenarios can have it as low as half the Higgs mass \cite{Carpenter:2020fnh}. As such, we include one benchmark, B1, with a sneutrino LSP. Other benchmarks feature a Higgsino LSP.  For all of these scenarios we choose we take all Dirac gaugino masses to be 3 TeV with negative sign choice for phase in regards to mass splitting terms.  We also choose $\lambda_S$ and $\lambda_T$ to be their $\mathcal{N}=2$ values $\frac{g^{'}}{\sqrt{2}}$ and $\frac{g^{'}}{\sqrt{2}}$. 

\begin{table}[]
    \centering
    \begin{tabular}{|c|c|c|c|}
         \hline 
        $\lambda_S$ & $\lambda_T$ & $M_{D1}=M_{D2}=M_{D3}$ & $m_{\chi}$ \\
        \hline
        $\frac{g'}{\sqrt{2}}$ & $\frac{g}{\sqrt{2}}$ & 3 TeV & 120 GeV \\
        \hline
    \end{tabular} \\[3ex]
    \begin{tabular}{|c||c|c|c|c|}
        \hline
         & $m_{\Tilde{\nu}}\ \mathrm{(GeV)}$ & $m_{\Tilde{\tau}}\ \mathrm{(GeV)}$ & $m_{\Tilde{t}_R}\ \mathrm{(GeV)}$ & $m_{\Tilde{t}_L}\ \mathrm{(GeV)}$ \\
         \hline 
        B1 & 100 & 150 & 800 & 900 \\
        \hline 
        B2 & 160 & 200 & 800 & 900 \\
        \hline 
        B3 & 500 & 550 & 1500 & 1000 \\
        \hline
        B4 & 500 & 550 & 800 & 900 \\
        \hline
    \end{tabular}
    \caption{The benchmark scenarios considered.}
    \label{benchmarks}
\end{table}

\section{Numerical Results}

Here we examine the production and decays of the scalar adjoints in the context of LHC searches. The Mathematica \cite{Mathematica} application Package-X \cite{Patel:2017px} was used to numerically evaluate the decay widths described above.  

\subsection{Branching Fractions}

\begin{figure}[h]
    \centering
    \includegraphics[width=0.7\linewidth]{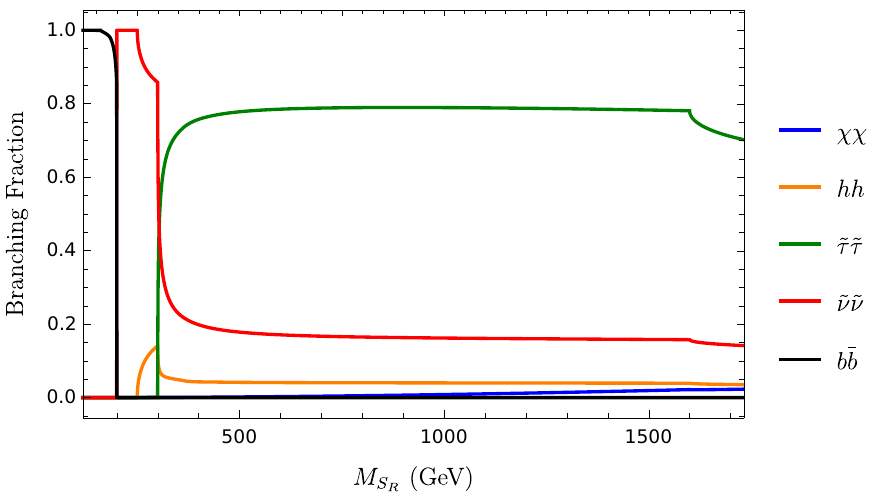} \\ \vspace{1cm}
    \includegraphics[width=0.7\linewidth]{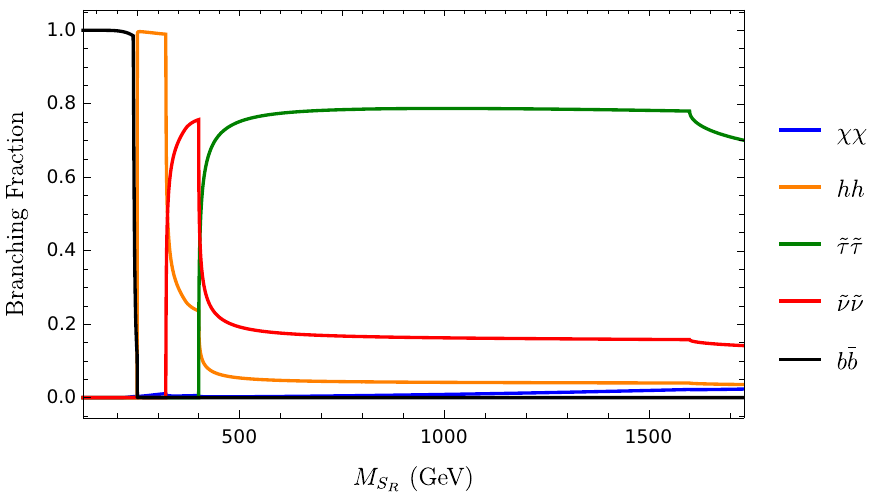} \\ \vspace{1cm}
    \includegraphics[width=0.7\linewidth]{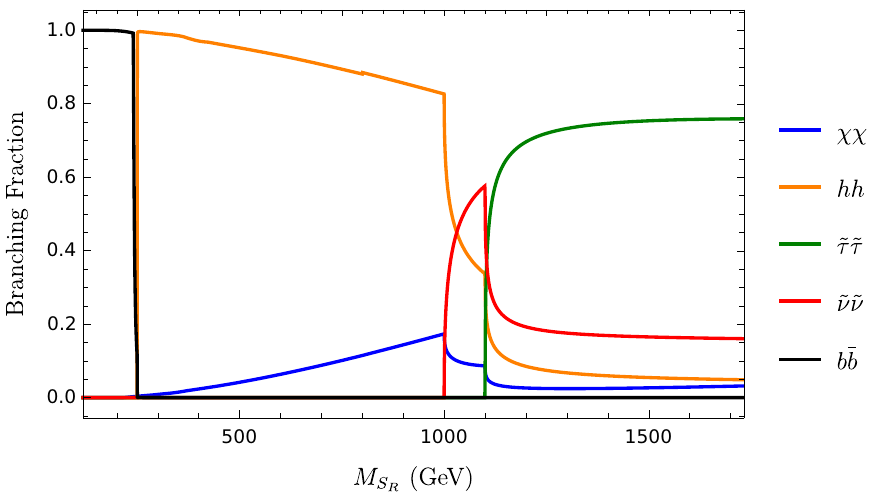}
    \caption{Branching fractions for decay of the scalar singlet in (top) benchmark B1, (middle) benchmark B2, and (bottom) benchmark B3.}
    \label{scalar_bfs}
\end{figure}

First we turn our attention to the $CP$-even sbino $S_R$. We have plotted branching fractions for three benchmark points are shown in Figure \ref{scalar_bfs}.  Due to the supersoft trilinear terms we find that decays of the scalar to right handed sleptons are dominant whenever kinematically allowed due to the large hypercharge of the right handed states.  In our benchmarks we have chosen to set the masses of left and right handed staus equal. Thus, in the all benchmarks the heavy singlet will most likely decay to stau pairs which then decay to a pair of $\tau$ leptons plus missing energy through the process $\tilde{\tau}\rightarrow \tau \chi_1^0$ or else decay $\tilde{\tau}\rightarrow \nu \chi_{1}^\pm$. Through the trilinear terms we also see a significant coupling of the sbino to left handed sleptons and light Higgs pairs. In our benchmarks sneutrinos are a bit lighter than left handed staus, leaving a small window for sneutrinos to become the dominant decay channel once it is kinematically available. Below the threshold of twice the sneutrino mass the di-Higgs decay is dominant. We can see a wide window of dominant di-Higgs decays in benchmarks B2 and B3 with heavier sneutrino masses. We see in Benchmark B1 in the upper plot that when the sneutrinos are of similar mass as the light Higgs, the decays into Higgses are subdominant due to symmetry factors. Recall that the possible LSPs are Higgsinos or sneutrinos. Thus, in this small mass window the decay of the singlet is invisible, for sneutrino LSP itself, or decays to a taus and charginos which themselves decay to mass degenerate neutralinos. This is a relatively unconstrained decay channel. For very low mass singlets, below the di-Higgs threshold, the singlet branching fraction is dominated by decay to the heaviest available fermion, in this case $b\bar{b}$ in all three benchmarks. However in this region the real singlet also has a significant Higgs component that is likely disfavored by fitting the Higgs production and decay observables.

\begin{figure}[h]
    \centering
    \includegraphics[width=0.7\linewidth]{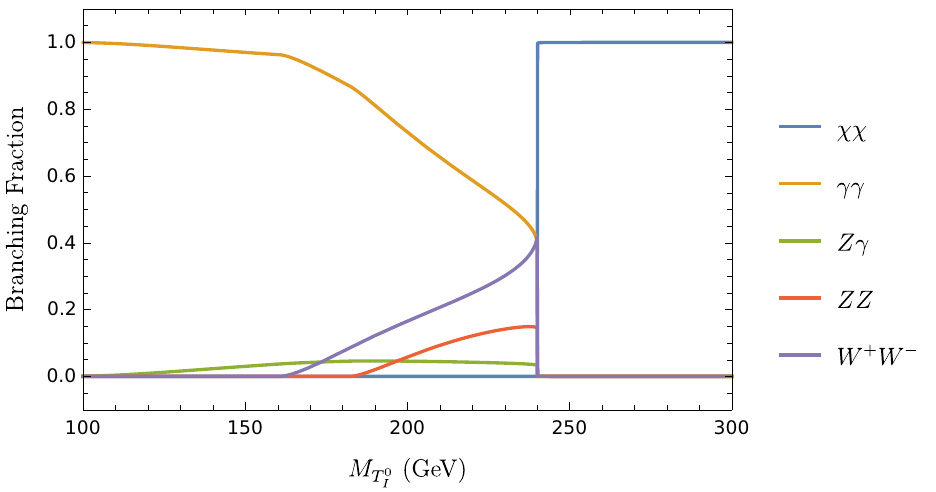}
    \caption{Branching fractions for decay of the pseudoscalar triplet in benchmark B2.}
    \label{pseudo-bfs-b2}
\end{figure}

Next, we consider the neutral $CP$-odd triplet swino. Representative branching fractions are shown in \hyperref[pseudo-bfs-b2]{Figure \ref{pseudo-bfs-b2}}. This particle  only has  significant couplings to gaugino or boson pairs.  Above twice the electroweak gaugino mass, the odd triplet decays to two neutralinos almost exclusively, meaning a singular production and decay would only show as missing energy. If the pseudoscalars are below twice the $\chi$ mass, they will predominantly decay to two photons. As we will discuss later, searches with di-photon resonance may be relevant for this model. There is an intermediate mass region in which the subdominant decay to $W$ pairs also makes up a non-trivial portion of the branching fraction.  

\begin{figure}[h]
    \centering
    \includegraphics[width=0.7\linewidth]{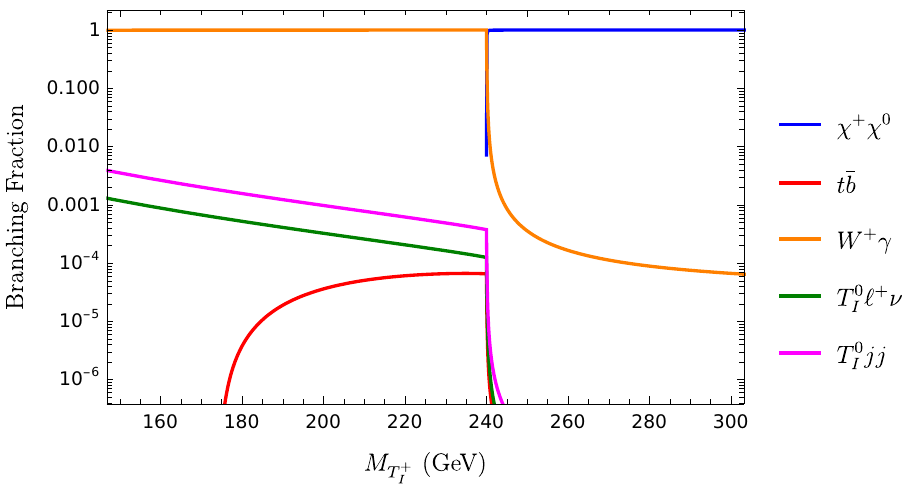}
    \caption{Branching fractions for decay of the odd charged triplet. Here the splitting between the charged and neutral state is fixed at 3 GeV, and all other parameters are as in benchmark B2.}
    \label{charged-bfs-b2}
\end{figure}

Finally, the branching fractions for the charged swino $T_I^+$ are shown in \hyperref[charged-bfs-b2]{Figure \ref{charged-bfs-b2}}. We see that heavy charged triplets favor decays into chargino-neutralino pairs. If the Higgsinos are the LSP this decay will be invisible due to the soft decay products of the chargino. If the sneutrino is the LSP the decay chain will contain a single tau as $\chi^{\pm}\chi^{0}\rightarrow \tilde{\nu}\tau\nu\tilde{\nu}$.  If the particle is too light to decay to Higgsinos, the dominant decay will be loop induced $W\gamma$ resonance.  This is an extremely interesting decay as it contains a hard photon and the decay products of the $W$, either a hard lepton or jets.

\subsection{Scalar Singlet Production Cross Sections and Constraints}

Here we evaluate the singlet's production cross section and compare to LHC searches. The package ManeParse \cite{Clark:2017mp} was used to integrate against CT10 parton distribution functions \cite{Lai:2010ct} and obtain the production cross section from gluon fusion. Explicitly, we have
\begin{equation}
    \sigma(pp \rightarrow S_R) = \frac{\pi^2}{M_{S_R}}\, \Gamma(S \to gg)\, \frac{1}{s} \int_{M_{S_R}^2/s}^1 \d x\, \frac{1}{x}\, f_g(x,M_{S_R}^2) f_g(M_{S_R}^2/sx,M_{S_R}^2),
\end{equation}
where $s$ is the center of mass energy of the event, and $f_g$ is the gluon distribution function. We only compute this production at leading order, and only including the lightest squarks, so we will enhance it with a $K$ factor of 2. 

\begin{figure}
    \centering
    \includegraphics[width=0.8\linewidth]{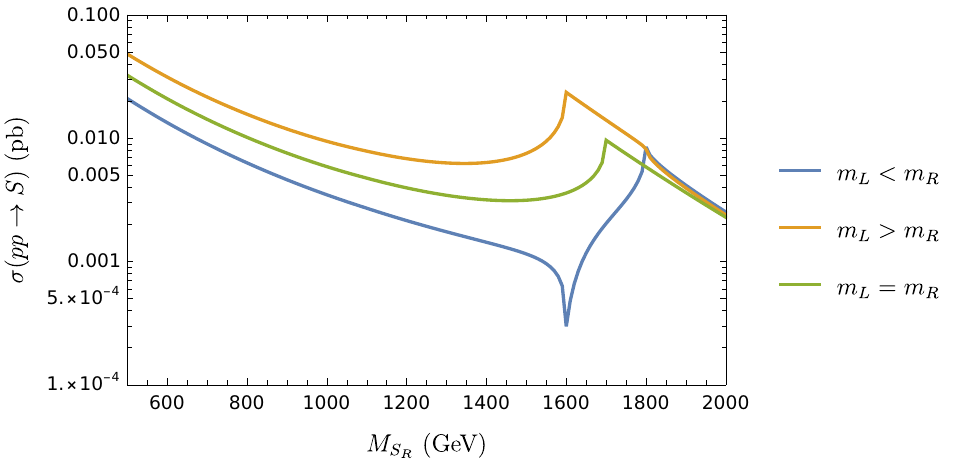}
    \caption{The scalar singlet production cross section via gluon fusion, in the case of (a) a lighter left stop, (b) a lighter right stop, and (c) degenerate left and right stops at 850 GeV, with all other parameters chosen as in benchmark B2.}
    \label{sigma-b2}
\end{figure}

As evident from \hyperref[sgwidth]{Equation \ref{sgwidth}}, there is a partial cancellation between contributions to $\sigma(pp \rightarrow S_R)$ from left- and right-chiral squark states due to the sign of hypercharge assignments. In particular, we find different qualitative behavior depending on whether the lightest squark is a left or right stop. \hyperref[sigma-b2]{Figure \ref{sigma-b2}} shows this explicitly.  Here we have shown the production cross section using three stop benchmarks, in blue a lighter left stop mass of 800 GeV and right stop mass 900 GeV; in green equal stop masses of 850 GeV; and in orange heavier left stop at 900 GeV and right stop at 800 GeV. When the left stop is lighter, the production cross section actually attains a minimum at the mass threshold due to this cancellation. This can be seen as a trough in the production plot \hyperref[sigma-b2]{Figure \ref{sigma-b2}} at the onshell threshold of the twice the left handed stop mass. We also see that qualitative shape of the production curve changes when when the right handed stop is lighter. In particular the trough is no longer present once $m_L\ge m_R$.  The overall production cross section is also higher in this case as the hypercharge of the left handed squark is larger.  There is a resonance peak feature in the production cross section plots, occurring at twice the right handed stop mass threshold, which drastically increases the production cross sections. 

\begin{figure}
    \centering
    \includegraphics[width=0.8\linewidth]{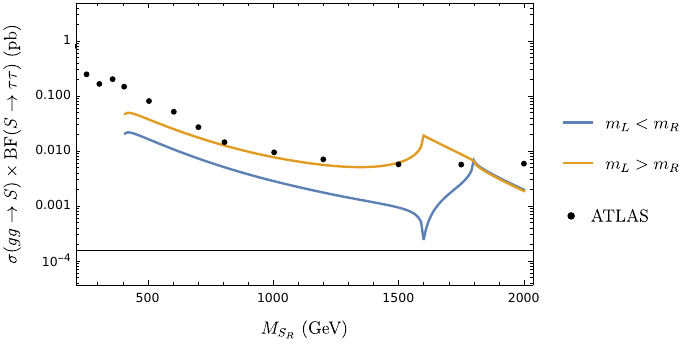}
    \caption{Resonant tau pair production in benchmark B2, assuming a 100\% branching ratio of the stau to tau plus missing energy. In black is the exclusion bound from \cite{Aaboud:2017sjh}.}
    \label{ditau_b2}
\end{figure}

\begin{figure}
    \centering
    \includegraphics[width=0.5\linewidth]{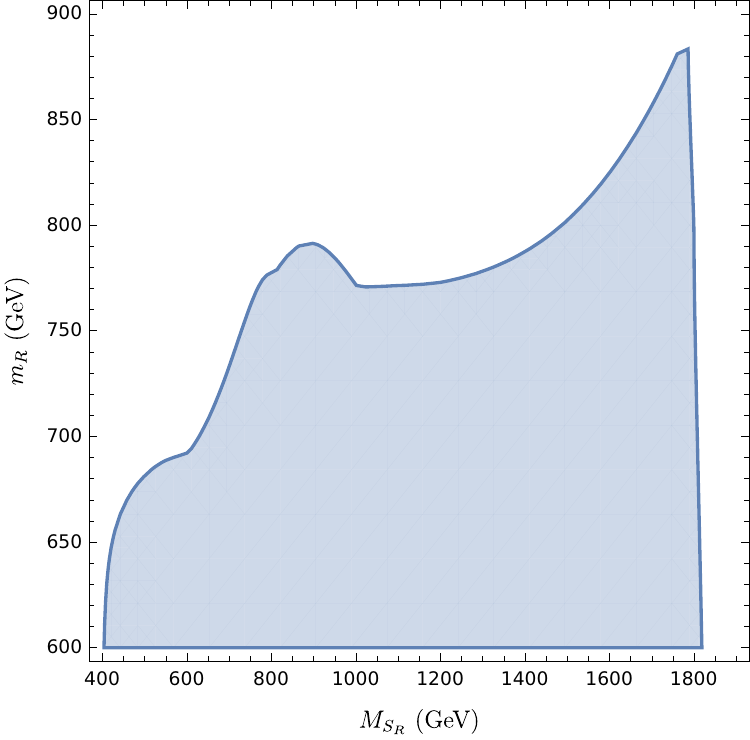}
    \caption{A plot of parameter space excluded by the ATLAS search in \cite{Aaboud:2017sjh}. Here the heavier (left) stop mass is fixed at 900 GeV, and all other parameters are as in benchmark B2.}
    \label{exclusion}
\end{figure}

For the scalar singlet, Figure \ref{ditau_b2} shows the stau pair production cross section via the singlet scalar in benchmark B2 for both stop mass orderings for lighter stop mass 800 GeV and heavier stop mass 900 GeV. Also shown for comparison (in black) are cross section exclusion bounds from an ATLAS search for scalar ditau resonances \cite{Aaboud:2017sjh}. This search reconstructed a binned transverse resonance using the transverse mass of tau decay products; thus we expect it to be sensitive to scalar decay scenario. Benchmark exclusions are very sensitive to stop masses, as seen in Figure \ref{ditau_b2}. We find that, in the case with a lighter right stop, there are regions of parameter space where the predicted cross section exceeds the bounds from the ditau search.  By contrast, the scenario with a lighter left stop has a lower production cross section and as such is far less constrained. \hyperref[exclusion]{Figure \ref{exclusion}} shows a slice of the $(M_{S_R},m_{\tilde{t}_R})$ parameter space where the production cross section is above the bound (assuming the left stop mass is heavier and fixed). For the case $m_L>m_R$, we investigate the exclusion from this search as follows. We use \textsc{Mathematica} to interpolate the excluded cross section limits from \cite{Aaboud:2017sjh}. We also compute numerically the predicted cross section as above, scanning over the singlet mass and $m_R$. Where the prediction exceeds the empirical bound, we mark the point as excluded. Results are shown in Figure \ref{exclusion}. We expect that future inclusive ditau searches will likely have the power to exclude much more parameter space.

\begin{figure}
    \centering
    \includegraphics[width=0.8\linewidth]{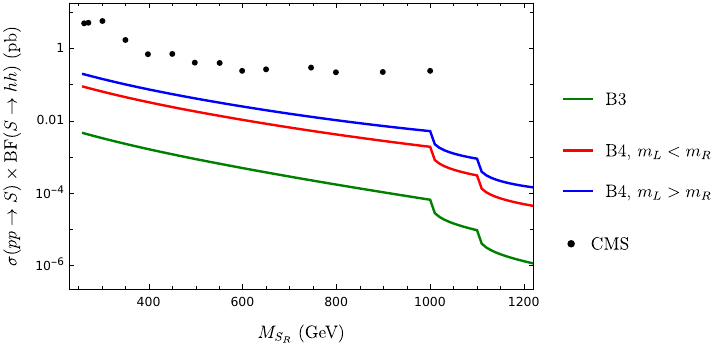}
    \caption{Resonant Higgs pair production in benchmarks B3 and B4 (shown with both stop mass orderings). Black points represent the CMS observed exclusion bound from \cite{Sirunyan_2020}.}
    \label{dihiggsb4}
\end{figure}

\begin{figure}
    \centering
    \includegraphics[width=0.8\linewidth]{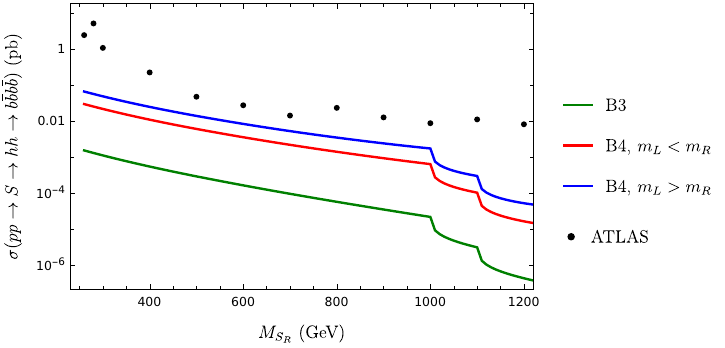}
    \caption{Resonant Higgs pair production and decay to bottom pairs in benchmarks B3 and B4 (shown with both stop mass orderings). Black points represent the ATLAS observed exclusion bound from \cite{ATLAS:2018rnh}.}
    \label{hh4b}
\end{figure}

For benchmarks such as B3 and B4, decays to Higgs pairs are the most significant channel for a wide range of parameter space. Even so, the production cross section falls well within constraints from Higgs pair production. In \hyperref[dihiggsb4]{Figure \ref{dihiggsb4}} we show scalar singlet production cross sections in benchmark 4, with the red and blue curves showing different choices of heavy and light stop, and the black dots showing observed CMS limits from a di-Higgs search \cite{Sirunyan_2020}.  The CMS search is not yet sensitive to this channel and misses by a factor of a few in production cross section. In addition, we may consider specific decay channels of the produced Higgs bosons, using the known branching ratios of the SM Higgs \cite{LHCHiggsCrossSectionWorkingGroup:2016ypw}. In Figure \ref{hh4b}, we compare this model to an ATLAS search for scalar resonances in the $b\bar{b}b\bar{b}$ final state \cite{ATLAS:2018rnh}. We again find that these benchmarks evade the constraints, though this is still the most promising channel. The branching ratios of the singlet to photons and jets are too small to be constrained by current dijet or diphoton searches (eg. \cite{ATLAS:2019itm,ATLAS:2017ayi}, as well as others referenced in \cite{Bechtle:2020pkv}).

\subsection{Charged Triplet Production Cross Sections}

\begin{figure}
    \centering
    \begin{tikzpicture}
        \begin{feynman}
            \vertex (a);
            \vertex [above left=of a](i1){\(q\)};
            \vertex [below left=of a](i2){\(q'\)};
            \vertex [right=2cm of a](b);
            \vertex [above right=of b](f1){\(T_I^+\)};
            \vertex [below right=of b](f2){\(T_I^0\)};
            
            \diagram*{
                (i1) --[fermion] (a),
                (i2) --[anti fermion] (a),
                (a) --[boson,edge label=\(W^+\)] (b),
                (b) --[scalar] (f1),
                (b) --[scalar] (f2),
            };
        \end{feynman}
    \end{tikzpicture} \ \ \ \ 
    \begin{tikzpicture}
        \begin{feynman}
            \vertex (a);
            \vertex [above left=of a](i1){\(q\)};
            \vertex [below left=of a](i2){\(q\)};
            \vertex [right=2cm of a](b);
            \vertex [above right=of b](f1){\(T_I^+\)};
            \vertex [below right=of b](f2){\(T_I^-\)};
            
            \diagram*{
                (i1) --[fermion] (a),
                (i2) --[anti fermion] (a),
                (a) --[boson,edge label=\(\gamma/Z\)] (b),
                (b) --[scalar] (f1),
                (b) --[scalar] (f2),
            };
        \end{feynman}
    \end{tikzpicture}
    \caption{Pair production of SU(2) adjoint scalars.}
    \label{pair-prod-diagram}
\end{figure}
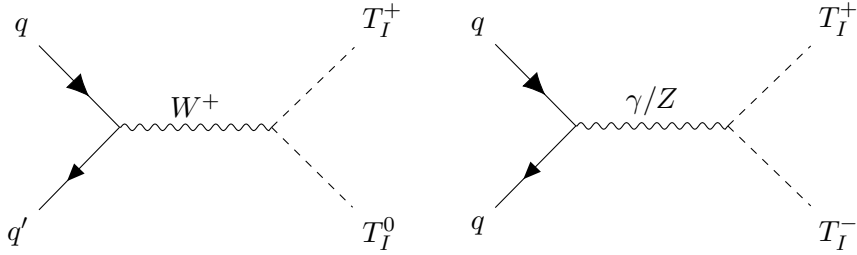

\begin{figure}
    \centering
    \includegraphics[width=0.8\linewidth]{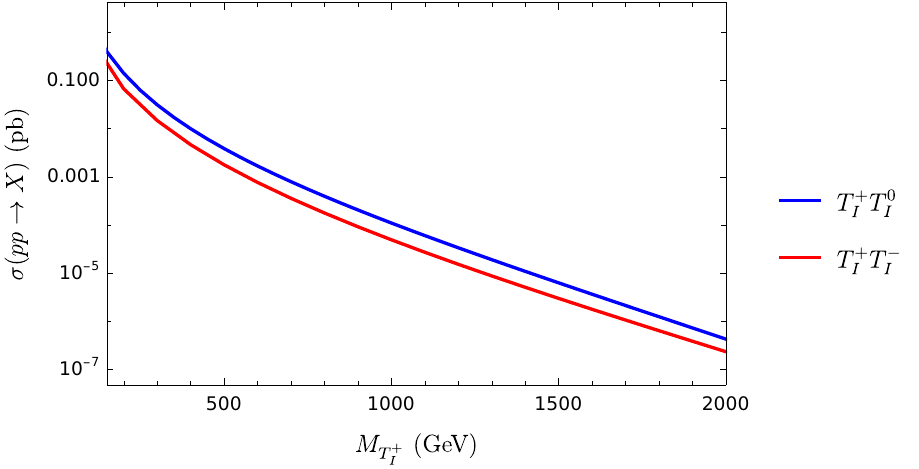}
    \caption{Pair production of triplets from $pp$ collisions at $\sqrt{s}= 13$ TeV.}
    \label{pair-production-plot}
\end{figure}

We will next turn to the pair production of scalar adjoints from electroweak processes. Due to their quantum numbers, the $CP$-odd neutral states $S_I$ and $T_I^0$ have no coupling to $\gamma$ or $Z$, so they cannot be pair produced, but there is leading order production of $T_I^\pm T_I^\pm$ pairs and $T_I^0 T_I^\pm$ pairs as shown in \hyperref[pair-prod-diagram]{Figure \ref{pair-prod-diagram}}. This process is very similar to that of slepton pair production in general SUSY models, which has been well-studied analytically and numerically \cite{Baer:1997nh,Bozzi:2004qq,Fiaschi:2019zgh}. Explicitly, we can express the total cross section to leading order as 
\begin{equation}
    \sigma = \int_{(M_{T_I^0}+M_{T_I^\pm})^2}^s \dd Q^2\ \frac{Q^2}{s} \int_{\frac{Q^2}{s}}^1 \frac{\dd x}{x} \sum_{i,j} f_i(x)f_j\left(\frac{Q^2}{xs}\right) \frac{\dd \hat{\sigma}_{ij}}{\dd Q^2},
\end{equation}
where the parton-level differential cross sections are modified from \cite{Baer:1997nh} only by the different quantum numbers for the triplet compared to the doublet sleptons:
\begin{gather}
    \frac{\dd \hat{\sigma}}{\dd Q^2}\left(q_i \bar{q}_j \rightarrow T_I^+ T_I^0\right) = \frac{2g^4}{576\pi} \beta^3 \frac{1}{(Q^2-M_W^2)^2}, \\[2ex]
    \begin{split}
    \frac{\dd \hat{\sigma}}{\dd Q^2}\left(q_j \bar{q}_j \rightarrow T_I^+ T_I^-\right) = \frac{g^4}{576\pi} \frac{\beta^3}{8c_W^4} \left[\frac{4c_W^4 (L_j^2 + R_j^2)}{(Q^2-M_Z^2)^2} + \frac{32c_W^4 e^4 e_j^2}{g^4 Q^4} + \right. \\[1.5ex]
    \left. + \frac{16c_W^4 e^2e_j(L_j+R_j)}{g^2}\frac{1}{Q^2(Q^2-M_Z^2)} \right],
    \end{split} \\[2ex]
    L_j = 2(I_j - e_j s_W^2),\ \ R_j = -2e_j s_W^2.
\end{gather}
Production of both $T_I^+ T_I^0$ and $T_I^+ T_I^-$ is shown in \hyperref[pair-production-plot]{Figure \ref{pair-production-plot}}. We see that production occurs between 1-100 fb for weak scale states. As expected $T_I^+ T_I^0$ production has the highest cross section. The final state will vary depending on if a decay to Higgsinos is accessible. If so, the decays of the adjoints will be invisible, and if not they will likely contain hard photons. In the case of heavy triplets with each decaying to invisible Higgsino pairs $TT\rightarrow \chi\chi\chi\chi$, it is likely the only limits will come from production of triplets along with an initial or final state gauge boson. Thus, mono jet (gluon) and mono vector boson searches used to constrain dark matter production are likely the only handles on this production process. In the case that Higgsino decays are inaccessible, the charged triplet will have large decay width into $W\gamma$ and the neutral triplet into $\gamma\gamma$. Thus the dominant process will be $pp\rightarrow T_I^+ T_I^0\rightarrow \gamma\gamma\gamma W^+$. This may be detected with a multi-photon search.

\section{Conclusion}

Non-minimal supersymmetric models will remain vitally important to both theory and experiment as we enter the next decade. $R$-symmetric models are particularly well-motivated, and in this work we have emphasized several interesting features of these models. The full scalar sector has rich phenomenology, and while the color-octet sgluons have stronger couplings to hadrons, the weak adjoints are worthy of examining in their own right. Nonzero vacuum expectation values of the neutral adjoint scalars can have a notable impact on the particle spectrum.  We have catalogued the possible decay processes of the weak adjoints in several benchmark scenarios. We have considered multiple benchmarks, focusing on scenarios with relatively light staus and stops, and have found a wide array of decays. These include significant invisible decay widths in the case of heavy Higgsinos, di-Higgs decays, and decays into interesting diboson resonances like $W\gamma$. 

As we have showed, there is a certain set of benchmarks already excluded by existing data, but parameter space remains relatively wide open, and there is much that can be done in future analyses. In particular, electroweak production of pseudoscalar triplets offers the possibility of multi-photon final states, and a dedicated search may well constrain these particles. Updates to di-Higgs and ditau searches will be more sensitive to the parameter space of $CP$-even scalar singlets. In addition it is possible that pseudoscalar triplet production may be investigated in vector boson fusion and associated production channels.

\section{Acknowledgements}

This work was funded in part by the United States Department of Energy under grant DE-SC0011726.

\bibliographystyle{Packages/JHEP}
\bibliography{bib/bibliography.bib}

\end{document}